\documentclass[journal=jpclcd,manuscript=letter,layout=twocolumn]{achemso}

\usepackage[version=3]{mhchem} 
\usepackage{xcolor}
\usepackage{hyperref}
\usepackage{verbatim}
\usepackage{amsmath}
\usepackage{titlecaps}
\usepackage{makecell}
\usepackage{tablefootnote}
\usepackage{diagbox}

\author{Ananth Govind Rajan}
\affiliation[IISC]{Department of Chemical Engineering, Indian Institute of Science, Bengaluru, Karnataka 560012, India}
\email{ananthgr@iisc.ac.in}

\title[An \textsf{achemso} demo]
  {Resolving the Debate Between Boltzmann and Gibbs Entropy: Relative Energy Window Eliminates Thermodynamic Inconsistencies and Allows Negative Absolute Temperatures}

\begin{document}

\begin{tocentry}
\includegraphics[width=5.08cm,height=5.08cm]{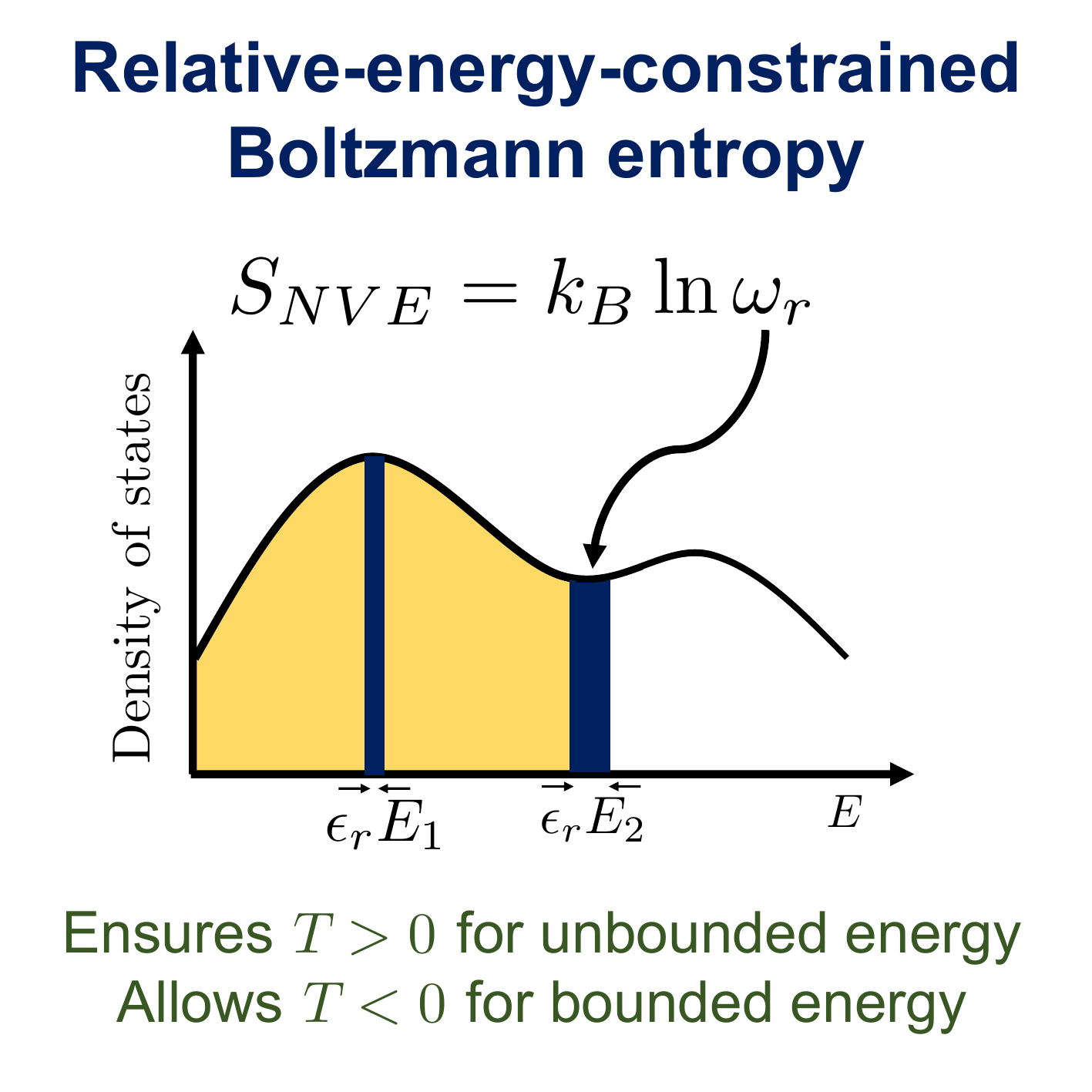}
\end{tocentry}

\begin{abstract}
Small systems consisting of a few particles are increasingly technologically relevant. In such systems, an intense debate in microcanonical statistical mechanics has been about the correctness of Boltzmann's surface entropy versus Gibbs' volume entropy. Both entropies have shortcomings---while Boltzmann entropy predicts unphysical negative/infinite absolute temperatures for small systems with unbounded energy, Gibbs entropy entirely disallows negative absolute temperatures, in disagreement with experiments. We consider a relative energy window, motivated by the Heisenberg energy-time uncertainty principle and eigenstate thermalization in quantum mechanics. The proposed entropy ensures positive, finite temperatures for systems without a maximum limit on their energy and allows negative absolute temperatures in bounded energy systems, e.g., with population inversion.  It also closely matches canonical ensemble predictions for prototypical systems, thus correctly describing the zero-point energy of an isolated quantum harmonic oscillator. Overall, we enable accurate thermodynamic models for isolated systems with few degrees of freedom.
\end{abstract}


The microcanonical ensemble in statistical mechanics, originally proposed by Gibbs over a century ago \cite{gibbs2014elementary}, continues to receive a significant extent of attention \cite{pearson1985laplace,uline2008generalized,davis2011calculation,Dunkel2014,corti2023microcanonical,GovindRajan2024}. In the physics literature, there exist two competing entropy definitions in the microcanonical ensemble: Gibbs’ volume entropy (also referred to as the Hertz entropy) \cite{gibbs2014elementary,Hertz1910,Rugh2001,Campisi2010,Campisi2015}, based on the phase space volume and Boltzmann’s surface entropy \cite{Gross2005,Frenkel2015} (also referred to as the Boltzmann-Planck entropy), based on the phase space density. Even though Boltzmann's surface entropy dominated textbook discourses \cite{Huang1987,Pathria2011,tuckerman2010statistical,Swendsen2012,shell2015thermodynamics}, many studies favored the use of the Gibbs entropy \cite{Berdichevsky1991,Rugh2001,Dunkel2006,Campisi2010,Campisi2015,Dunkel2014}. Note that the terms Gibbs entropy for the definition based on the phase space volume and Boltzmann entropy for that based on the phase space density are based on the prevalent names in the literature, although Gibbs' seminal work did acknowledge the possibility of both entropy definitions\cite{gibbs2014elementary}. 

About a decade ago, Dunkel and Hilbert indicated that Boltzmann’s surface entropy leads to violations of equipartition and unphysical results, i.e., negative/infinite temperatures, for some model systems with unbounded spectra \cite{Dunkel2014}, i.e., those without a maximum limit on the allowable energies. Subsequently, several articles appeared that pointed out that Gibbs entropy suffers from serious objections, such as the consideration of all states with energy lower than energy $E$, rather than only the states at the desired energy and the inability to allow for negative absolute temperatures in systems with non-monotonically increasing density of states \cite{Schneider2014,Frenkel2015,Swendsen2015,Swendsen2017,Lavis2019,Lustig2019,corti2023microcanonical}. Physically, a negative absolute temperature implies that the entropy of a system reduces as the energy increases. This is indeed possible in systems with bounded energy spectra as higher energies would involve more particles occupying the highest-energy level, thus reducing the entropy with an increase in energy.

Although the Boltzmann and Gibbs entropies \textit{usually} agree for very large systems, they differ significantly and lead to widely varying temperatures for isolated systems with few degrees of freedom \cite{Barbatti2022}. (In some cases, even in the thermodynamic limit, e.g., the case of a two-level system (see below), they can disagree.)  In this regard, the principles of statistical mechanics can be applied to systems with few degrees of freedom, as the statistical averaging is carried out over the classical phase space or the available quantum states, rather than over the degrees of freedom \cite{Jensen1985,Chirikov1973,Cerino2014,corti2023microcanonical}. Thus, even quantities like the temperature of an isolated quantum state \cite{Lipka-Bartosik2023,Mitchison2022,Burke2023b} or the entropy of a few-body system are well defined \cite{Phillies2000,Santos2011}.

In light of this fact, that both entropy definitions have shortcomings for small systems \cite{Swendsen2017} is particularly unnerving. Indeed, the lack of a universal microcanonical entropy formulation not only adversely impacts the fundamental physical understanding of the thermodynamics of small systems \cite{Andersen2001}, but also could impede technological progress in the future, wherein systems of few atoms, molecules, or electrons are increasingly becoming physically realizable \cite{Hartman2018}. Not only that, the applicability of statistical mechanics to small systems is particularly important in molecular biology\cite{Puglisi2018}, nuclear physics,\cite{Chomaz2006,Borderie2019} and astronomical systems\cite{Gross2006}. In fact, Puglisi et al. have discussed that the conceptual framework of statistical mechanics can successfully predict the behavior of small systems.\cite{Puglisi2017} 

In this Letter, we show that the consideration of a relative energy tolerance (also called the energy window) in the microcanonical ensemble eliminates all thermodynamic inconsistencies put forth in previous work pertaining to systems with unbounded spectra while at the same time allowing negative absolute temperatures in systems with bounded spectra. In addition, in the Supporting Information, we show that a consistency condition mentioned in ref. ~\cite{Dunkel2014} is not valid even for the Gibbs entropy, thus invalidating the former's use to support the latter. Now, by the term unbounded, we refer to the fact that the maximum allowable system energy is infinite, i.e., it has no upper bound. Conversely, the term bounded means that the system energy has an upper limit, as seen in the case of a two-level spin system later in the paper. Note that the term ``bounded'' does not have any reference to the energy constraint employed in Boltzmann-type entropies.

It is also important to clarify that our focus is not on the entropy \textit{formulae} ascribed to Boltzmann and Gibbs, which are both equally valid. Rather, we seek to examine the controversy surrounding the correct choice of the microstates that are counted to determine the entropy in Boltzmann's entropy formula, which is applicable to a microcanonical system as
\begin{equation}
S=k_B \ln w
\end{equation}
where $S$ denotes the entropy of the system, $k_B$ is the Boltzmann constant, and $w$ is the number of microstates, each having equal probability $1/w$. In the above formula, Gibbs' entropy uses the phase space volume ($\Omega$) for $w$, and Boltzmann's entropy uses the phase space density ($\omega$) for it. Note that Gibbs' entropy is unrelated to the Gibbs entropy formula, which allows one to calculate entropy in any statistical-mechanical ensemble as:
\begin{equation}
    S= -k_B\sum_i p_i \ln p_i
\end{equation}
where $p_i$ denotes the probability of occurrence of the microstate $i$, the summation running over all possible microstates corresponding to the considered macrostate.

\begin{figure}[h]
\includegraphics[width=9cm]{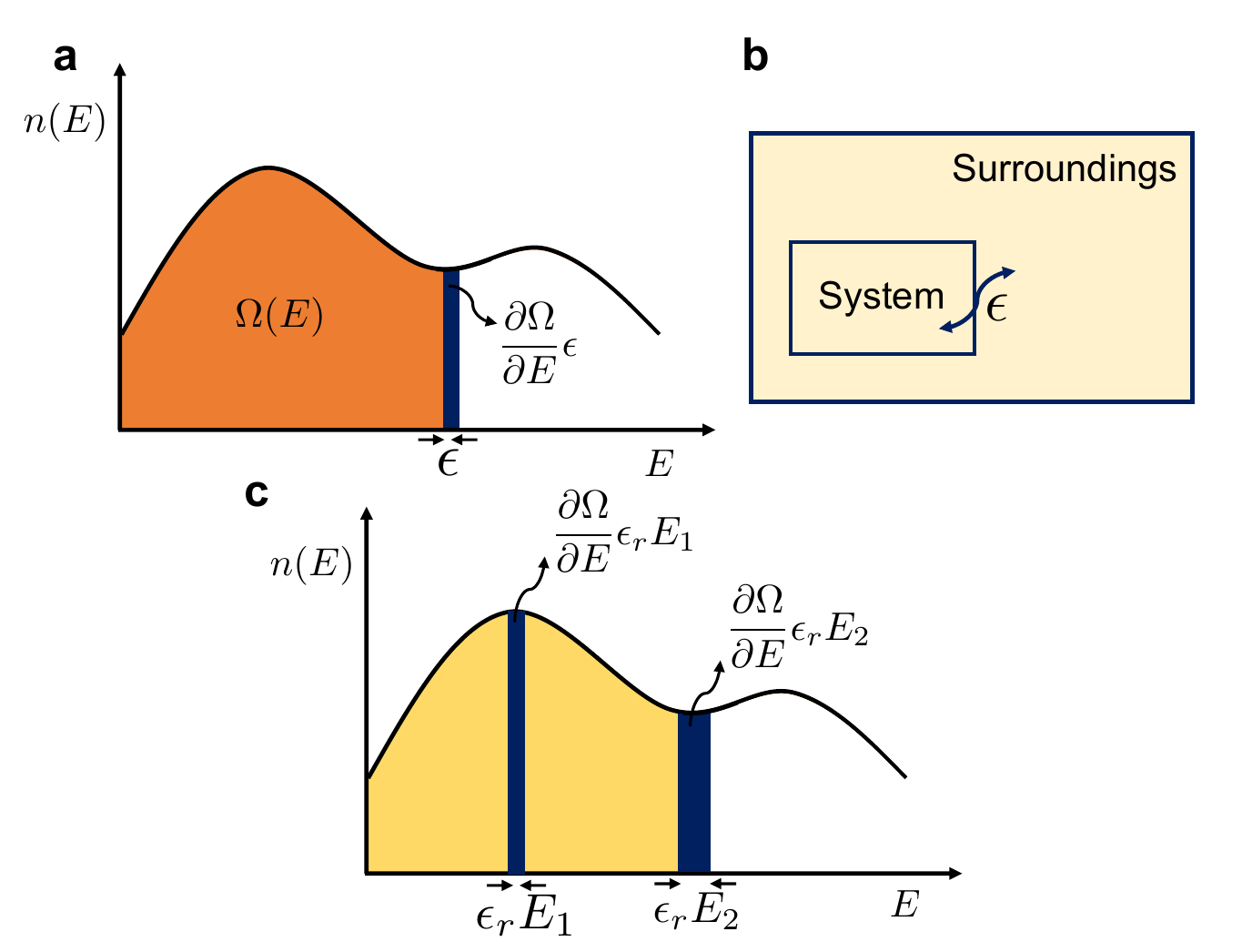}
\caption{Motivation and explanation of the relative-energy-constrained Boltzmann entropy. (a) For a density of states $n(E)$, the Gibbs entropy is defined by the logarithm of the area under the curve, i.e., $\Omega$, while the Boltzmann entropy is defined by the logarithm of the change in area due to a fixed energy perturbation $\epsilon$, i.e., $\frac{\partial\Omega}{\partial E}\epsilon$. (b) Interaction of a system with its surroundings leads to smearing of the system's energy $E$ by a perturbation $\epsilon$. (c) The proposed entropy considers an energy-dependent energy perturbation, given as $\epsilon_rE$.}
\label{fgr:defn}
\end{figure}

Now, considering the microcanonical ($NVE$) ensemble with fixed number of particles $N$, volume $V$, and energy $E$, the Gibbs entropy ($S_G$) is defined as $S_G=k_B\ln\Omega$, where $\Omega$ denotes the volume of the classical phase space or the number of quantum states in the region $E'<E$ (Figure \ref{fgr:defn}a), and $E'$ denotes the possible values the system's energy can take. For a classical system, this can be written as
\begin{equation}
\label{eq:cmTheta}
    \Omega = \frac{1}{h^{dN}N!}\int d\textbf{$\mathrm{r}$}^N d\textbf{$\mathrm{p}$}^N \Theta(E-H)
\end{equation}
where $\Theta$ is the Heaviside step function, $\Theta(x)=\begin{cases}
    1; \;\;\; x>0\\
    0; \;\;\; x\le 0
\end{cases}$, $h$ is Planck's constant, $d$ is the dimensionality of the system, $H$ is the Hamiltonian of the system, $\mathrm{\textbf{r}}^N$ denotes the positions of the particles, and $\mathrm{\textbf{p}}^N$ denotes their momenta. Alternatively, for a quantum-mechanical system, we can write
\begin{equation}
\label{eq:qmTheta}
        \Omega = \mathrm{Tr}[\Theta(E-H)]
\end{equation}
where $\mathrm{Tr}$ denotes the trace, i.e., the sum over quantum states in a Hilbert space. The ensemble corresponding to this entropy definition has sometimes been referred to as the uniform ensemble, where the probability of every state with $H\le E$ is equal,\cite{Tolman1938,tuckerman2010statistical} rather than only those with $H = E$ as in the microcanonical ensemble. In contrast, the Boltzmann entropy ($S_B$) is conventionally defined as $S_B=k_B\ln\omega$, where
\begin{equation}
\label{eq:cmDelta}
    \omega = \frac{1}{h^{dN}N!}\int d\textbf{$\mathrm{r}$}^N d\textbf{$\mathrm{p}$}^N \delta\left(\frac{H-E}{\epsilon}\right) 
\end{equation}
or
\begin{equation}    
\label{eq:qmDelta}
\omega = \mathrm{Tr}\left[\delta\left(\frac{H-E}{\epsilon}\right)\right]
\end{equation}
and $\epsilon$ is an absolute energy tolerance, as visualized in Figure \ref{fgr:defn}a. 

Physically, the tolerance appears because no system is perfectly isolated, and some amount of perturbation or broadening in the energy occurs due to the system's interactions with its surroundings (Figure \ref{fgr:defn}b), leading to an ``energy window''. See, e.g., Pathria\cite{Pathria2011}, who states that ``almost every system has some contact with its surroundings, however little it may be; as a result, its energy cannot be defined sharply.'' In his work, Gibbs did not comment on the origin of the energy tolerance, simply stating that a Gaussian spread of energy could be considered.\cite{gibbs2014elementary} However, Mayer and Mayer, as well as Landau and Lifshitz provided physical intuition into the energy tolerance using quantum mechanics, whereby Heisenberg's uncertainty principle requires that $\sigma_E \Delta t \ge \frac{\hbar}{2}$, indicating that if $\sigma_E$, the uncertainty in energy equals zero, the system would have to remain isolated in the considered energy level for an infinite amount of time, i.e., $\Delta t \rightarrow \infty$.\cite{Mayer1940,landau2013statistical} In real-world scenarios, however, since the observational time period, i.e., $\Delta t$, is finite, $\sigma_E$ also needs to be finite. Note that, in this work, $\Delta E$ represents the difference in energy between two consecutive eigenstates, whereas $\sigma_E$ represents the uncertainty in energy of the system, i.e., the energy window.

\begin{figure}[h]
\includegraphics[width=7cm]{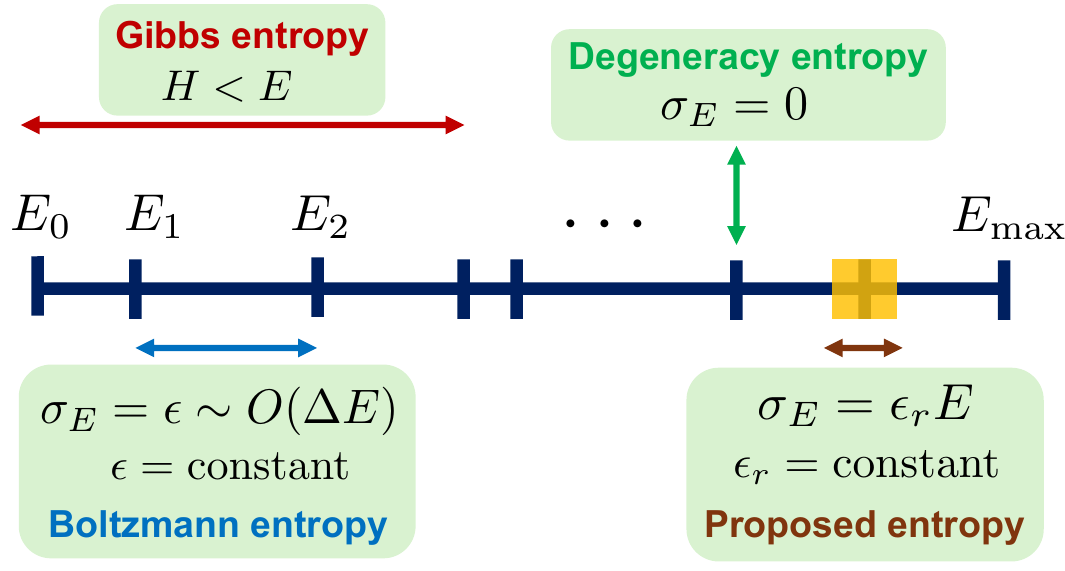}
\caption{Difference between the proposed, Boltzmann, Gibbs, and degeneracy entropies in terms of which energy levels are included in the calculation of the microcanonical partition function, $w$, denoted using double-ended arrows. $E_0$ denotes the ground-state energy, $E_1$ the first excited-state energy, and so on. $E_{\max}$ denotes the maximum-allowable energy, which is a finite number for a bounded spectrum, and infinity for an unbounded spectrum. States with energy less than or equal to $E$ are allowed by Gibbs entropy, states with a fixed tolerance $\epsilon$ around $E$, typically of the order of $\Delta E$ are allowed by Boltzmann entropy, states only at energy $E$ are included in $w$ for the degeneracy entropy, and states with a relative-energy tolerance $\epsilon_r E$ around $E$ are included in the proposed entropy.}
\label{fgr:defn2}
\end{figure}

From Eqs. (\ref{eq:cmTheta})-(\ref{eq:qmTheta}) and (\ref{eq:cmDelta})-(\ref{eq:qmDelta}), one can infer that
\begin{eqnarray}
    \omega=\epsilon \frac{\partial\Omega}{\partial E}
    \label{eq:DOS}
\end{eqnarray}
Even for a discrete spectrum, although direct enumeration is often used as a simplification \cite{Park2022}, Eq. (\ref{eq:DOS}) is preferred \cite{Pathria2011,landau2013statistical}, as it allows the correct definition of entropy for nondegenerate quantum systems \cite{Burke2023,Gurarie2007}. Indeed, we argue that entropy should account for the uncertainty of a state not only in terms of any degeneracies but also in terms of any perturbations in energy due to the quantum-mechanical restriction on knowing the energy of a state exactly over finite periods of time. Thus, we make a distinction between ``degeneracy entropy'', which only accounts for the uncertainty in a microstate due to several microstates possessing exactly the same energy, and the Boltzmann/Gibbs/proposed entropies, which also include the uncertainty due to the allowable spread in the energy, as shown in Figure \ref{fgr:defn2}. We show that the degeneracy entropy fails to capture the correct thermodynamic behavior even for most discrete systems, except in the thermodynamic limit in one case.

Typically, $\epsilon$ has been interpreted to be a fixed but small constant in Eq. (\ref{eq:DOS}), and this has led to inconsistent results for simple systems with unbounded energy levels, e.g., negative temperatures for an ideal gas with one degree of freedom or infinite temperature for a quantum harmonic oscillator and an ideal gas with two degrees of freedom. We suggest here that instead of using an absolute energy tolerance, the energy tolerance in the microcanonical ensemble should be defined relative to the system's energy (Figure \ref{fgr:defn}c). Indeed, a system with 1 J of energy and with 1 eV = 1.602 $\times$ 10\textsuperscript{-19} J of energy should not use the same tolerance in the microcanonical ensemble. This is because the higher the energy of a system, there exists a possibility of greater interaction with its surroundings (as acknowledged by Landau and Lifshitz\cite{landau2013statistical}) or higher levels of internal perturbations modifying the energy of the system from the calculated energy levels (as remarked by Mayer and Mayer\cite{Mayer1940}), causing more broadening or perturbation in the system's energy. 

Now, in principle, a single eigenstate (stationary state) of a system exhibits zero uncertainty in energy, i.e., $\sigma_E = 0$. However, this is only when it is undisturbed for an infinite amount of time. In practice, the system's wavefunction would necessarily be entangled with that of its surroundings, involving some level of thermalization. The thermalization time of a quantum system goes as $\frac{h}{E_{\alpha}-E_{\beta}}$ where $\alpha$ and $\beta$ denote different eigenstates, as the eigenstates involved dephase over this timescale.\cite{Srednicki1994} Although an upper limit for the thermalization time of many-body, interacting, and \textit{perfectly isolated} systems is the Heisenberg time, i.e., when $E_{\alpha}-E_{\beta}$ equals the level spacing, $\Delta E$,\cite{Lezama2021} a lower limit for small, noninteracting, and weakly coupled systems could be obtained by considering $E_{\alpha}$ as the eigenstate closest to the mean energy of the system ($E$) and $E_{\beta}$ as the ground state, so that $\Delta t \sim \frac{h}{E}$. For this lower limit of thermalization time, which is inversely proportional to the system's energy, the uncertainty in energy would be proportional to the system's energy, as per Heisenberg's uncertainty principle.

Therefore, we propose that the constraint inside the delta function should be expressed in terms of the relative error of $H_r=\frac{H}{E}$ from 1. In terms of the phase space integrals or Hilbert space summations, doing so amounts to applying a constraint on $H$ without using up a degree of freedom. Hence, classically, one should write
\begin{equation}
\label{eq:newentropy1}
\omega_r = \frac{1}{h^{dN}N!}\int d\textbf{r}^N d\textbf{p}^N \delta\left(\frac{H_r-1}{\epsilon_r}\right) 
\end{equation}
whereas for a quantum-mechanical system, we should write
\begin{equation}
\label{eq:newentropy2}
        \omega_r = \text{Tr}\left[\delta\left(\frac{H_r-1}{\epsilon_r}\right)\right]
\end{equation}
where $\epsilon_r$ is a relative error tolerance, and is a very small (dimensionless) number independent of $E$. Now, even though $\epsilon_r$ affords larger deviations of $H$ from $E$, these deviations are still arbitrarily small due to the former's small value. Further, as per the above discussion, $\epsilon_r E$ represents the uncertainty in knowing the energy of the system exactly, i.e., $\sigma_E$, even for a single eigenstate, due to the quantum-mechanical uncertainties introduced by the environment (see Figure \ref{fgr:defn2}). Thus, $\epsilon_r E=\sigma_E$ can be lower than even the energy difference between two successive eigenstates ($\Delta E$), unlike what has been assumed previously ($\sigma_E \sim O(\Delta E$)).\cite{Rigol2008}

Accordingly, in Eqs. (\ref{eq:newentropy1}) and (\ref{eq:newentropy2}), we use the subscript $r$ to denote the use of the relative deviation from the desired energy. It can be shown that 
\begin{equation}
    \omega_r = \epsilon_r E \frac{\partial\Omega}{\partial E}
\end{equation}
as indicated in Figure \ref{fgr:defn}c, in the high-temperature limit, when the states form a continuum. At low temperatures/energies, however, the discreteness of the energy levels implies that
\begin{equation}
    \omega_r = \epsilon_r E \frac{\Delta\Omega}{\Delta E}
\end{equation}
where $\Delta\Omega$ represents the change in the number of states as the energy changes from the present eigenstate to the immediate next one by $\Delta E$. Thus, $\Delta\Omega = 1$ for a non-degenerate state, whereas $\Delta \Omega>1$ for degenerate states. In this work, we consider examples of both cases---the single quantum harmonic oscillator and 1D quantum particle in a box are examples where all energy levels are non-degenerate, whereas the two-level system involves degenerate energy levels. In all the examples, we use a \textit{forward difference} to evaluate $\Delta E$ and $\Delta \Omega$ and found thermodynamically consistent results. We note that apart from issues in defining temperature using the Boltzmann entropy, there are also problems related to the chemical potential, as discussed by Corti et al.\cite{corti2023microcanonical} They remarked that a backward difference is more suitable to evaluate the chemical potential for small systems using the original Boltzmann entropy in the microcanonical ensemble.\cite{corti2023microcanonical} Future work can therefore explore the use of backward differences to evaluate $E$ vs. $T$ relationships, as well as both forward/backward differences for evaluating the chemical potential, using the proposed entropy definition.

We note that there is some precedent in the literature for adopting an energy-dependent tolerance \cite{Gurarie2007,Burke2023}. However, in both these studies, the energy tolerance was assumed to be proportional to $\sqrt{C_CT_C^2}$, where $C_C = \frac{\partial E}{\partial T_C}$ is the canonical heat capacity of the system and $T_C$ denotes its canonical temperature. Thus, the energy tolerance depends upon thermodynamic variables ($T_C$ and $C_C$) not directly defined in the micrononical ensemble, leading to a forced equivalence with the canonical ensemble. In fact, in the Supporting Information, we show that the energy-dependent energy tolerance proposed in refs. \cite{Burke2023} and \cite{Gurarie2007} leads to the incorrect equation of state for an isolated quantum harmonic oscillator, unlike the energy tolerance definition proposed in this work. It is also noteworthy that the equations proposed here do not change the definition of the density of states, which is still $\frac{\partial\Omega}{\partial E}$, such that the Laplace transform-based relationship between the canonical partition function and the density of states still holds\cite{corti2023microcanonical}. Similarly, the phase space density is unchanged (as it is multiplied everywhere in the phase space by the fixed energy $E$, a constant factor that disappears upon normalization), such that Liouville's theorem still holds in a manner similar to that for the original Boltzmann entropy.

We now apply the new approach for calculating the microcanonical relative-energy-constrained Boltzmann entropy, $S_{B_r} = k_B\ln \omega_r$, and the corresponding temperature as either
\begin{equation}
\label{eq:Tbr}
    \frac{1}{T_{B_r}} = \frac{\partial S_{B_r}}{\partial E}
\end{equation}
or
\begin{equation}
\label{eq:Tbr2}
    \frac{1}{T_{B_r}} = \frac{\Delta S_{B_r}}{\Delta E}
\end{equation}
to several example systems. One may think that converting an absolute energy constraint to a relative energy constraint would not affect the calculation of entropy significantly. However, as seen in Table \ref{tbl:comparison}, we conclusively show that using a relative constraint preserves the degrees of freedom of a system and thus leads to thermodynamically consistent behavior for various isolated systems with few degrees of freedom.\\
\\
\begin{table*}
\begin{minipage}{\textwidth}
\caption{Comparison of the predictions of the degeneracy, Gibbs, Boltzmann, and proposed entropies in the microcanonical ensemble with the corresponding predictions in the canonical ensemble.}
\label{tbl:comparison}
\resizebox{\textwidth}{!}{\begin{tabular}{ | c | c | c | c | c | c | }
\hline
 \backslashbox{\textbf{System $\downarrow$}}{\textbf{Entropy $\rightarrow$}} & \textbf{Degeneracy} & \textbf{Gibbs} & \textbf{Boltzmann} & \textbf{Proposed} & \textbf{Canonical} \\ 
 \hline
 \makecell{3D ideal gas\\(high $T$)} & not defined & $E=\frac{3N}{2}k_BT_G$ & $E=\left(\frac{3N}{2}-1\right)k_BT_B$ & $E=\frac{3N}{2}k_BT_{B_r}$ & $E=\frac{3N}{2}k_BT_C$    \\  
\hline
\makecell{Harmonic oscillator\\ (high $T$)} & $T_D\rightarrow \infty$ & $E=k_BT_G-\frac{h\nu}{2}$ & $T_B \rightarrow \infty$ & $E = k_BT_{B_r}$ & $E=k_BT_C$ \\
\hline
\makecell{Harmonic oscillator\\ (any $T$)\footnote{Note that $E_0=\frac{1}{2}h\nu$ and $\theta=\frac{h\nu}{k_B}$ for the harmonic oscillator.}} & $T_D\rightarrow \infty$ & $\frac{E-E_0}{h\nu}=\frac{3-\exp\left(\frac{\theta}{T_G}\right)}{2\exp\left(\frac{\theta}{T_G}\right)-2}$ & $T_B \rightarrow \infty$ & $\frac{E-E_0}{h\nu}=\frac{1}{\exp\left(\frac{\theta}{T_{B_r}}\right)-1}$ & $\frac{E-E_0}{h\nu}=\frac{1}{\exp\left(\frac{\theta}{T_C}\right)-1}$ \\
\hline
\makecell{Particle in\\ a 1D box\\(high $T$)} & $T_D\rightarrow \infty$ & $E=\frac{k_BT_G}{2}$ & $E=-\frac{k_BT_B}{2}$ & $E=\frac{k_BT_{B_r}}{2}$ & $E=\frac{k_BT_C}{2}$\\
\hline
\makecell{Two-level\\system\footnote{$N_0$ and $N_1$ denote, respectively, the number of particles in the ground and excited state, respectively.}\\(any $T$)}& $T_D = \frac{E_1}{k_B\ln\left(\frac{N_0}{N_1+1}\right)}$ & $T_G=\frac{E_1}{k_B\ln\left(\frac{\sum_{N_0'=1}^{N_1+1} \frac{N!}{N_0'!N_1'!}}{\sum_{N_0'=1}^{N_1} \frac{N!}{N_0'!N_1'!}}\right)}$ &  $T_B  = \frac{E_1}{k_B\ln\left(\frac{N_0-1}{N_1+2}\right)}$ & $T_{B_r}  =  \frac{E_1}{k_B\ln\left(\frac{(N_1+1)(N_0-1)}{N_1(N_1+2)}\right)}$ & $T_C=\frac{E_1}{k_B\ln\left(\frac{N_0}{N_1}\right)}$\\
\hline
\makecell{Negative\\ temperatures?} & Yes & No & Yes & Yes & Yes\\
 \hline
\end{tabular}}
\end{minipage}
\end{table*}

First, we consider a $d$-dimensional monatomic ideal gas consisting of $N$ atoms of mass $m$ each, occupying a total volume $V$. This system has unbounded energy levels (since there is no upper limit on the classical kinetic energy of a particle) and thus should not have negative absolute temperature, as correctly argued by Dunkel and Hilbert\cite{Dunkel2014}. The volume of the phase space for this system is \cite{Dunkel2006}
\begin{equation}
\Omega = \frac{V^N}{h^{dN}N!}\frac{(2\pi m)^{\frac{dN}{2}}}{\Gamma\left(\frac{dN}{2}+1\right)}E^{\frac{dN}{2}}.
\end{equation}
Further, the number of states can be written as 
\begin{equation}
\omega = \frac{V^N}{h^{dN}N!}\frac{(2\pi m)^{\frac{dN}{2}}}{\Gamma\left(\frac{dN}{2}\right)}E^{\frac{dN}{2}-1}\epsilon.
\end{equation}
Note that, for this system, the degeneracy entropy is not defined as the continuum of states implies an uncountable number of precise states. Nevertheless, the Gibbs temperature ($T_G$) can be calculated using the equation $\frac{1}{T_G}=\frac{\partial S_G}{\partial E}$ to obtain $E=\frac{dN}{2}k_B T_G$. However, using the analogous equation for Boltzmann temperature ($T_B$), assuming a \textit{fixed energy tolerance} $\epsilon$, leads to $E=\left(\frac{dN}{2}-1\right)k_B T_B$. Note that if $dN=1$, i.e., the system has one degree of freedom, one obtains negative temperatures, whereas if $dN=2$, i.e., the system has two degrees of freedom, one gets infinite temperature \cite{Dunkel2014}. (In these cases as well, the center of mass velocity is zero as configurations, wherein the particle moves in both positive and negative directions along a given translational degree of freedom, are included in the phase space average.) This is a serious shortcoming of the Boltzmann entropy, as pointed out by Dunkel and Hilbert \cite{Dunkel2014}, as one is led to predict negative/infinite temperature for a system whose energy is unbounded. Considering a \textit{fractional energy tolerance}, one however obtains $\omega_r = \frac{V^N}{h^{dN}N!}\frac{(2\pi m)^{\frac{dN}{2}}}{\Gamma\left(\frac{dN}{2}\right)}E^{\frac{dN}{2}}\epsilon_r$, which combined with Eq. (\ref{eq:Tbr}) leads to
\begin{equation}
E=\frac{dN}{2}k_B T_{B_r}.
\end{equation}
The above equation not only disallows negative/infinite temperatures for an ideal gas but fixes the disagreement between the energy of a three-dimensional ideal gas as traditionally calculated in the canonical $\left(E=\frac{3N}{2}k_BT\right)$ and microcanonical $\left(E=\left(\frac{3N}{2}-1\right)k_BT\right)$ ensembles (see Table \ref{tbl:comparison})! Although there is no requirement in statistical mechanics for the predictions from various ensembles to match\cite{hill1994thermodynamics}, similar predictions of the canonical and microcanonical ensemble fall out of the analysis done here using the relative-energy-constrained Boltzmann entropy. This may not always be the case, as shown later in this work for the case of a two-level system. Nevertheless, the relative-energy-constrained Boltzmann entropy eliminates the possibility of negative/infinite temperatures for systems in which there is no maximum limit on the allowed energies. Moreover, the proposed entropy fixes the disagreement between the ideal gas temperature based on the inverse kinetic energy as obtained from the microcanonical ensemble and that from the zeroth law of thermodynamics.\cite{corti2023microcanonical}\\

Next, for a quantum harmonic oscillator, the energy of the system is given as \cite{griffiths2018introduction} $E_n=h \nu\left(n+\frac{1}{2}\right)$, where $n=0,1,2,...$ and $\nu$ represents the frequency of vibration of the oscillator. It follows that $\Omega=1+n=\frac{E}{h\nu}+\frac{1}{2}$. Since $0\le n < \infty$, the system's energy is unbounded. At the high-temperature limit, the states can be treated as a continuum, so that $k_BT_G=\frac{h\nu}{2}+E$. Although applying the original definition of Boltzmann entropy leads to infinite temperature since $\omega=\frac{\epsilon}{h\nu}$, the newly proposed definition (Eq. (\ref{eq:newentropy2})) results in $\omega_r=\frac{\epsilon_r E}{h\nu}$, leading to $E=k_BT_{B_r}$, which predicts finite, positive temperatures. Again, only the relative-energy-constrained Boltzmann entropy leads to predictions consistent with the canonical ensemble (Table \ref{tbl:comparison})! Note that the degeneracy entropy is zero for every energy level because $\Delta\Omega=1$, indicating that the degeneracy entropy would predict $T_D\rightarrow\infty$, where $T_D$ denotes the temperature corresponding to the degeneracy entropy, similar to the prediction of the Boltzmann entropy in this case.

\begin{figure}[h]
\includegraphics[width=9cm]{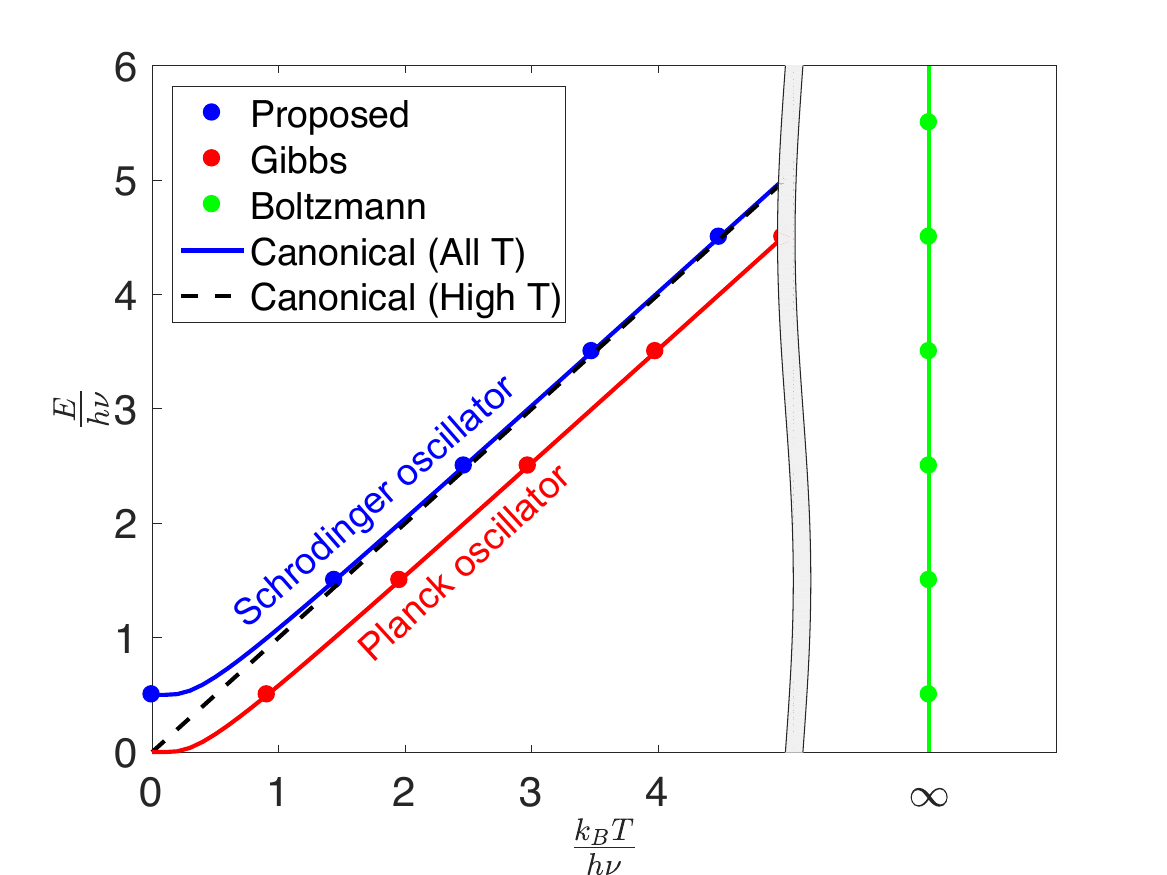}

\caption{Comparison of various microcanonical entropies in terms of their ability to predict the thermodynamic behavior of a quantum harmonic oscillator. Blue, red, and green circles represent predictions of the proposed, Gibbs, and Boltzmann/degeneracy entropies in the microcanonical ensemble. The canonical prediction is shown using a blue line and the high-temperature canonical prediction is shown using a dashed black line. In these cases, lines are simply a guide. Only the proposed entropy predicts the existence of the correct zero-point energy at 0 K and degenerates to the limit of $k_BT$ at high temperatures.}
\label{fgr:comparison}
\end{figure}

One can also derive expressions valid for the temperature under all cases, using discretized definitions of the partition function and temperature (see, e.g., ref. ~\cite{Miranda2013}), as shown in the Supporting Information and in Table \ref{tbl:comparison}. The ensuing $E$ vs. $T$ relationships for various entropy choices are plotted in Figure \ref{fgr:comparison}. Therein, one sees that only the \textit{relative-energy-constrained} Boltzmann entropy predicts thermodynamically consistent results at \textit{all} temperatures, including the existence of the zero-point energy $\left(\frac{1}{2}h\nu\right)$ and agreeing with the high-temperature prediction of $E=k_BT$. On the contrary, while the Gibbs entropy fails to predict the existence of a zero-point energy (i.e., leads to a Planck oscillator rather than a Schrodinger oscillator; see Figure \ref{fgr:comparison} and ref. ~\cite{Pathria2011} for a definition of Planck and Schrodinger oscillators), the Boltzmann entropy predicts infinite temperatures for all energies. Note that, to the best of the author's knowledge, this is the first derivation of the temperature of a \textit{single, isolated} quantum harmonic oscillator, i.e., in the microcanonical ensemble; previous expressions relied on the consideration of multiple oscillators, in the limit of a large number of oscillators (see, e.g., refs. \cite{Shirts2021} and \cite{kardar2007statistical}). Note that, as seen in Figure \ref{fgr:comparison}, in the microcanonical ensemble, only discrete values of $E$, i.e., $E_0+\frac{1}{2}h\nu$ (where $E_0$ is the zero-point energy) are allowed (as per the spectrum of the harmonic oscillator) as indicated by the blue circles, whereas, in the canonical ensemble, all possible values of $E$ are allowed due to the contact with a thermal reservoir, as depicted by the continuous blue line. \\

Now, we consider a quantum particle in a one-dimensional (1D) box. The energy of this system is given as \cite{griffiths2018introduction} $E_n=\frac{an^2}{L^2}$, where $a=\frac{\hbar^2\pi^2}{2m}$ and $n=1,2,3,...$, so that $\Omega=L\sqrt{\frac{E}{a}}$ and consequently $\omega = \frac{\partial\Omega}{\partial E}\epsilon = \frac{L\epsilon}{2\sqrt{Ea}}$. Since there is no limit on how large $n$ can be, the system has an unbounded energy spectrum. At high temperatures, the states form a continuum, and we find that $E=\frac{k_BT_G}{2}$ and $P_G=T_G\frac{\partial S_G}{\partial L} = \frac{2E}{L}$. This statistical-mechanical pressure equals the thermodynamic pressure calculated as $P_T=-\frac{\partial E}{\partial L}$. Now, although the conventional Boltzmann entropy leads to a negative temperature $\left(E=-\frac{k_BT_B}{2}\right)$ and negative pressure $\left(P_B=-\frac{2E}{L}\right)$, the newly proposed definition of Boltzmann entropy fixes these predictions to yield $\omega_r = \frac{L\epsilon_r}{2}\sqrt{\frac{E}{a}}$, $P_{B_r}=\frac{2E}{L}$, and $E=\frac{k_BT_{B_r}}{2}$, as seen in Table \ref{tbl:comparison}. Again, because $\Delta\Omega=1$, the degeneracy entropy predicts infinite temperature for every energy level (i.e., $T_D\rightarrow\infty$), thus not being useful as a definition of entropy.

We now demonstrate that the newly proposed definition of Boltzmann entropy still allows for negative temperatures for systems having non-monotonic density of states \cite{Swendsen2015,Swendsen2017,Frenkel2015}, i.e., a bounded phase space. Considering the well-known case of a two-level system, with energy levels 0 and $E_1$ and $N_0$ particles occupying the ground state and $N_1$ particles occupying the excited state, the energy of the system is $E=N_1E_1$ and the number of particles is $N=N_0+N_1$. This system has an upper bound on its energy equal to $NE_1$. It follows that the number of states with energy less than or equal to $E$ is
\begin{equation}
    \Omega = \sum_{N_0'=1}^{N_1} \frac{N!}{N_0'!N_1'!} 
\end{equation}
where $N_1=E/E_1$ and $N_1'=N-N_0'$. We use the discrete definitions of the partition function and temperature so that the derivation holds for both low and high temperatures. Accordingly, the Gibbs temperature is obtained as
\begin{eqnarray}
    T_G = \frac{\Delta E}{\Delta S_G} = \frac{E_1}{k_B\ln\left(\frac{\sum_{N_0'=1}^{N_1+1} \frac{N!}{N_0'!N_1'!}}{\sum_{N_0'=1}^{N_1} \frac{N!}{N_0'!N_1'!}}\right)}.
\end{eqnarray}
Moving further, the original Boltzmann partition function is given as
\begin{eqnarray*}
    \omega= \frac{\Delta\Omega}{\Delta E}\epsilon  =  \frac{N!}{(N_0-1)!(N_1+1)!}\frac{\epsilon}{E_1}.
\end{eqnarray*}
where $\Delta \Omega$ is obtained as the difference in the summation expressions of $\Omega$ with $N_0'$ going up to $(N_1+1)$ and $N_1$. Thus
\begin{eqnarray*}
    S_B= k_B\ln \left(\frac{N!}{(N_0-1)!(N_1+1)!}\frac{\epsilon}{E_1}\right).
\end{eqnarray*}
For fixed $N$, the absolute-energy-constrained Boltzmann temperature can be determined to be
\begin{equation}
    T_B  = \frac{\Delta E}{\Delta S_B} = \frac{E_1}{k_B\ln\left(\frac{N_0-1}{N_1+2}\right)}.
\end{equation}
For a large number of particles, the temperature can simply be written as $T_B = \frac{E_1}{k_B\ln\left(\frac{N_0}{N_1}\right)}$. Now, if $E_1>0$ and $N_0<N_1$, i.e., we consider a system exhibiting ``population inversion'', we find that $T_B<0$. Analogously, the relative-energy-constrained Boltzmann partition function is given as
\begin{eqnarray*}
    \omega_r = \frac{\Delta\Omega}{\Delta E}\epsilon_r E = \frac{N!}{(N_0-1)!(N_1+1)!}\frac{\epsilon_r E}{E_1}.
\end{eqnarray*}
This leads to the following expression for the temperature:

\begin{equation}
    T_{B_r}  = \frac{\Delta E}{\Delta S_{B_r}} = \frac{E_1}{k_B\ln\left(\frac{(N_1+1)(N_0-1)}{N_1(N_1+2)}\right)},
\end{equation}
which also simplifies, for a large number of particles, to $T_{B_r} = \frac{E_1}{k_B\ln\left(\frac{N_0}{N_1}\right)}$,
also allowing negative temperatures. Note that the degeneracy entropy predicts $T = \frac{E_1}{k_B\ln\left(\frac{N_0}{N_1+1}\right)}$ as the number of degenerate states changes from $\frac{N!}{N_0!N_1!}$ to $\frac{N!}{(N_0-1)!(N_1+1)!}$ as the energy changes from $N_1E_1$ to $(N_1+1)E_1$.

\begin{figure}[h]
\includegraphics[width=9cm]{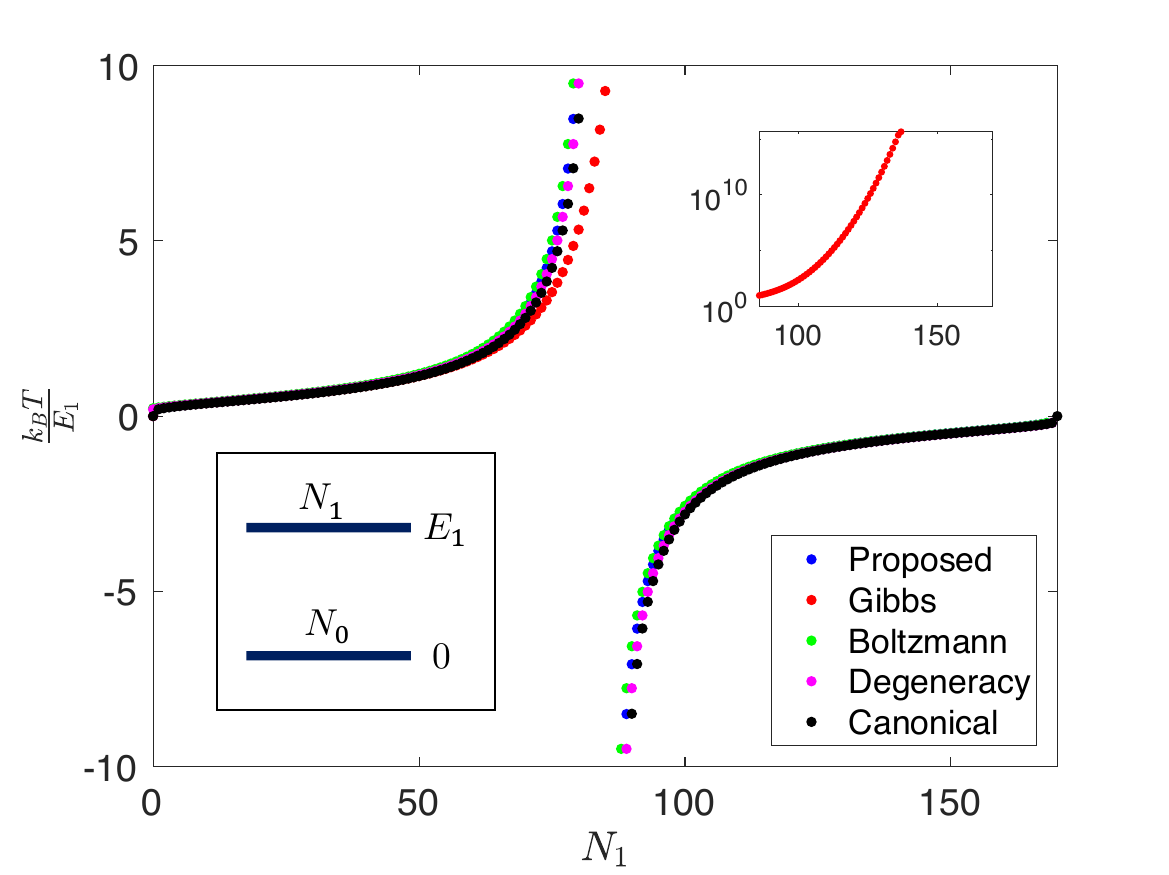}
\caption{Comparison of the predictions made by various microcanonical entropies for a two-level system consisting of $N=170$ particles. $E_1$ and $N_1$ denote, respectively, the energy and the number of particles in the excited state, and $N_0$ denotes the number of particles in the ground state, as shown in the bottom-left inset. The top-right inset shows the Gibbs entropy's temperature prediction on a log scale.}
\label{fgr:twolevel}
\end{figure}

Figure \ref{fgr:twolevel} shows the temperature predictions made by the proposed, Boltzmann, Gibbs, degeneracy, and canonical (see the Supporting Information) entropies, for a two-level system consisting of $N=170$ particles, where the predictions are obtained numerically. Although the factorials disappear in the temperature expressions for entropies other than the Gibbs entropy, as seen above, the factorials are retained in the Gibbs entropy expression. The chosen $N=170$ is thus the largest one for which $N!$ is calculable in double-precision floating-point format. In Figure \ref{fgr:twolevel}, one sees that the proposed entropy leads to a slightly lower temperature than the conventional Boltzmann entropy. Moreover, we see that Gibbs entropy does not admit negative temperatures at all due to a monotonically increasing total number of states. Instead, for system configurations wherein $T_B$ and $T_{B_r}$ are negative, $T_G$ is exceptionally large, which actually represents a cooler system since negative absolute temperatures are hotter than positive absolute temperatures.

Interestingly, while the predictions of the degeneracy entropy are closer to the original Boltzmann entropy, the predictions of the proposed entropy are closer to canonical entropy, considering that the entropy definitions are staggered due to the use of finite differences. These findings support the earlier claims that Gibbs entropy does not account for systems with bounded spectra in a thermodynamically consistent manner \cite{Ramsey1956,Vilar2014,Frenkel2015}. They also indicate that Gibbs and Boltzmann entropies can disagree, even in the thermodynamic limit. Moreover, these results provide an experimental route to test the validity of the proposed and Boltzmann entropies by measuring the temperature of a two-level system. 

Note that, according to the above discussion, the newly proposed relative-energy-constrained Boltzmann entropy ($S_{B_r}$) allows for both negative and positive temperatures, and in the appropriate contexts in which they should be seen, i.e., systems with bounded and unbounded phase space, respectively (see Table \ref{tbl:comparison}). Therefore, this work indicates that previous negative temperature measurements for various systems are indeed correct \cite{Purcell1951,Javan1959,Hakonen1994,Mosk2005,Rapp2010,Braun2013,Baudin2023}, unlike what is stated in ref. ~\cite{Dunkel2014}. In fact, another argument against negative temperature, viz., the incorrect interpretation of Carnot efficiencies being more than one has also been addressed by Abraham and Penrose \cite{Abraham2017}. The reader is also referred to the comprehensive review by Baldovin et al., which lays out the statistical mechanics of systems with negative absolute temperatures and discusses how they are consistent with both equilibrium and non-equilibrium thermodynamics.\cite{Baldovin2021}

To conclude, we introduced a relative energy constraint in the calculation of the microcanonical partition function, which eliminates negative/infinite temperatures for systems with unbounded spectra (i.e., without an upper limit on the allowed energy), such as ideal gases with a few degrees of freedom, a quantum harmonic oscillator, and a quantum particle confined in an infinite well. At the same time, the proposed entropy definition allows for negative temperatures for systems with non-monotonic density of states, i.e., a bounded spectrum (with an upper limit on energy), such as a two-level system with population inversion. This makes the proposed entropy definition the one that is the closest to predictions from the canonical ensemble in each case considered (see Table \ref{tbl:comparison}), although such equivalence was not imposed in this work and may not arise for all possible systems, as shown for the two-level system. We also explained how the degeneracy-based entropy is not applicable in many cases and also does not agree with canonical-ensemble predictions.

The physical insight underlying the proposed relative-energy-constrained entropy is that systems with higher energy will interact more with their surroundings, thus causing a larger perturbation in their energy. This proposal is motivated using the fact that the interactions of a system with its surroundings can never be fully eliminated, combined with the Heisenberg uncertainty principle and an eigenstate thermalization time inversely proportional to the system energy. Mathematically, while an absolute energy constraint uses up one degree of freedom from the system (by fixing the value of $E$), the use of a relative energy constraint fixes only the ratio $H/E$, thus retaining the original number of degrees of freedom of the system and resulting in the correct scaling of energy with temperature for various systems. We show that this is critical to describing correct physics, e.g., recovering the zero-point energy of a quantum harmonic oscillator in the microcanonical ensemble and avoiding unphysical temperatures. Nevertheless, we remark that, depending on the underlying physics of the system at hand, one would have to carefully make a choice between applying an absolute or a relative energy constraint. Another open question is determining the value of $\epsilon_r$ for a given microcanonical system very weakly coupled to a bath.

In summary, we proposed a resolution to the ongoing debate in microcanonical statistical mechanics on the use of the Gibbs volume entropy versus Boltzmann surface entropy. In this regard, we also proposed an experimental route to test the validity of the proposed and the Boltzmann entropies using a two-level system. Overall, this work sets the foundation for using the relative-energy-constrained Boltzmann surface entropy in the microcanonical ensemble of statistical mechanics. Therefore, the work puts on a firm theoretical footing the existence of negative temperatures for systems with bounded energy spectra. Finally, this work lays a solid foundation for accurate theoretical descriptions of isolated systems with few degrees of freedom using statistical mechanics.

\begin{acknowledgement}
A.G.R. acknowledges the Infosys Foundation, Bengaluru for an Infosys Young Investigator grant. A.G.R. thanks Profs. Subroto Mukerjee, Sumilan Banerjee, Sudeep N. Punnathanam, Binny J. Cherayil, K. Ganapathy Ayappa, Narendra M. Dixit, Prabal K. Maiti, and Prabhu R. Nott, and Dr. Garrett R. Dowdy  for insightful discussions, and Profs. Diptiman Sen, Ankur Gupta, and Sanjeev K. Gupta for valuable comments/suggestions regarding the manuscript.
\end{acknowledgement}

\begin{suppinfo}
Derivation of the energy expression for a quantum harmonic oscillator in the microcanonical ensemble valid at all temperatures, high-temperature limits for the energy of a quantum harmonic oscillator, examination of the energy-dependent energy window proposed in previous work, the two-level system in the canonical ensemble, and examining the consistency condition mentioned in previous work.
\end{suppinfo}

\bibliography{apssamp}

\providecommand{\noopsort}[1]{}\providecommand{\singleletter}[1]{#1}%
\providecommand{\latin}[1]{#1}
\makeatletter
\providecommand{\doi}
  {\begingroup\let\do\@makeother\dospecials
  \catcode`\{=1 \catcode`\}=2 \doi@aux}
\providecommand{\doi@aux}[1]{\endgroup\texttt{#1}}
\makeatother
\providecommand*\mcitethebibliography{\thebibliography}
\csname @ifundefined\endcsname{endmcitethebibliography}  {\let\endmcitethebibliography\endthebibliography}{}
\begin{mcitethebibliography}{67}
\providecommand*\natexlab[1]{#1}
\providecommand*\mciteSetBstSublistMode[1]{}
\providecommand*\mciteSetBstMaxWidthForm[2]{}
\providecommand*\mciteBstWouldAddEndPuncttrue
  {\def\EndOfBibitem{\unskip.}}
\providecommand*\mciteBstWouldAddEndPunctfalse
  {\let\EndOfBibitem\relax}
\providecommand*\mciteSetBstMidEndSepPunct[3]{}
\providecommand*\mciteSetBstSublistLabelBeginEnd[3]{}
\providecommand*\EndOfBibitem{}
\mciteSetBstSublistMode{f}
\mciteSetBstMaxWidthForm{subitem}{(\alph{mcitesubitemcount})}
\mciteSetBstSublistLabelBeginEnd
  {\mcitemaxwidthsubitemform\space}
  {\relax}
  {\relax}

\bibitem[Gibbs(1902)]{gibbs2014elementary}
Gibbs,~J.~W. \emph{Elementary Principles in Statistical Mechanics}; Charles Scribner's Sons, 1902\relax
\mciteBstWouldAddEndPuncttrue
\mciteSetBstMidEndSepPunct{\mcitedefaultmidpunct}
{\mcitedefaultendpunct}{\mcitedefaultseppunct}\relax
\EndOfBibitem
\bibitem[Pearson \latin{et~al.}(1985)Pearson, Halicioglu, and Tiller]{pearson1985laplace}
Pearson,~E.~M.; Halicioglu,~T.; Tiller,~W.~A. {Laplace-Transform Technique for Deriving Thermodynamic Equations from the Classical Microcanonical Ensemble}. \emph{Phys. Rev. A} \textbf{1985}, \emph{32}, 3030\relax
\mciteBstWouldAddEndPuncttrue
\mciteSetBstMidEndSepPunct{\mcitedefaultmidpunct}
{\mcitedefaultendpunct}{\mcitedefaultseppunct}\relax
\EndOfBibitem
\bibitem[Uline \latin{et~al.}(2008)Uline, Siderius, and Corti]{uline2008generalized}
Uline,~M.~J.; Siderius,~D.~W.; Corti,~D.~S. On the Generalized Equipartition Theorem in Molecular Dynamics Ensembles and the Microcanonical Thermodynamics of Small Systems. \emph{J. Chem. Phys.} \textbf{2008}, \emph{128}, 124301\relax
\mciteBstWouldAddEndPuncttrue
\mciteSetBstMidEndSepPunct{\mcitedefaultmidpunct}
{\mcitedefaultendpunct}{\mcitedefaultseppunct}\relax
\EndOfBibitem
\bibitem[Davis(2011)]{davis2011calculation}
Davis,~S. Calculation of Microcanonical Entropy Differences from Configurational Averages. \emph{Phys. Rev. E} \textbf{2011}, \emph{84}, 050101\relax
\mciteBstWouldAddEndPuncttrue
\mciteSetBstMidEndSepPunct{\mcitedefaultmidpunct}
{\mcitedefaultendpunct}{\mcitedefaultseppunct}\relax
\EndOfBibitem
\bibitem[Dunkel and Hilbert(2014)Dunkel, and Hilbert]{Dunkel2014}
Dunkel,~J.; Hilbert,~S. {Consistent Thermostatistics Forbids Negative Absolute Temperatures}. \emph{Nat. Phys.} \textbf{2014}, \emph{10}, 67--72\relax
\mciteBstWouldAddEndPuncttrue
\mciteSetBstMidEndSepPunct{\mcitedefaultmidpunct}
{\mcitedefaultendpunct}{\mcitedefaultseppunct}\relax
\EndOfBibitem
\bibitem[Corti \latin{et~al.}(2023)Corti, Ohadi, Fariello, and Uline]{corti2023microcanonical}
Corti,~D.~S.; Ohadi,~D.; Fariello,~R.; Uline,~M.~J. Microcanonical Thermodynamics of Small Ideal Gas Systems. \emph{J. Phys. Chem. B} \textbf{2023}, \emph{127}, 3431--3442\relax
\mciteBstWouldAddEndPuncttrue
\mciteSetBstMidEndSepPunct{\mcitedefaultmidpunct}
{\mcitedefaultendpunct}{\mcitedefaultseppunct}\relax
\EndOfBibitem
\bibitem[{Govind Rajan}(2024)]{GovindRajan2024}
{Govind Rajan},~A. {Pedagogical Approach to Microcanonical Statistical Mechanics via Consistency with the Combined First and Second Law of Thermodynamics for a Nonideal Fluid}. \emph{Journal of Chemical Education} \textbf{2024}, 2448--2457\relax
\mciteBstWouldAddEndPuncttrue
\mciteSetBstMidEndSepPunct{\mcitedefaultmidpunct}
{\mcitedefaultendpunct}{\mcitedefaultseppunct}\relax
\EndOfBibitem
\bibitem[Hertz(1910)]{Hertz1910}
Hertz,~P. {{\"{U}}ber die Mechanischen Grundlagen der Thermodynamik}. \emph{Ann. Phys. (Leipzig)} \textbf{1910}, \emph{338}, 225--274\relax
\mciteBstWouldAddEndPuncttrue
\mciteSetBstMidEndSepPunct{\mcitedefaultmidpunct}
{\mcitedefaultendpunct}{\mcitedefaultseppunct}\relax
\EndOfBibitem
\bibitem[Rugh(2001)]{Rugh2001}
Rugh,~H.~H. {Microthermodynamic Formalism}. \emph{Phys. Rev. E} \textbf{2001}, \emph{64}, 055101\relax
\mciteBstWouldAddEndPuncttrue
\mciteSetBstMidEndSepPunct{\mcitedefaultmidpunct}
{\mcitedefaultendpunct}{\mcitedefaultseppunct}\relax
\EndOfBibitem
\bibitem[Campisi and Kobe(2010)Campisi, and Kobe]{Campisi2010}
Campisi,~M.; Kobe,~D.~H. {Derivation of the Boltzmann principle}. \emph{Am. J. Phys.} \textbf{2010}, \emph{78}, 608--615\relax
\mciteBstWouldAddEndPuncttrue
\mciteSetBstMidEndSepPunct{\mcitedefaultmidpunct}
{\mcitedefaultendpunct}{\mcitedefaultseppunct}\relax
\EndOfBibitem
\bibitem[Campisi(2015)]{Campisi2015}
Campisi,~M. {Construction of Microcanonical Entropy on Thermodynamic Pillars}. \emph{Phys. Rev. E} \textbf{2015}, \emph{91}, 052147\relax
\mciteBstWouldAddEndPuncttrue
\mciteSetBstMidEndSepPunct{\mcitedefaultmidpunct}
{\mcitedefaultendpunct}{\mcitedefaultseppunct}\relax
\EndOfBibitem
\bibitem[Gross and Kenney(2005)Gross, and Kenney]{Gross2005}
Gross,~D.~H.; Kenney,~J.~F. {The Microcanonical Thermodynamics of Finite Systems: The Microscopic Origin of Condensation and Phase Separations, and the Conditions for Heat Flow from Lower to Higher Temperatures}. \emph{J. Chem. Phys.} \textbf{2005}, \emph{122}\relax
\mciteBstWouldAddEndPuncttrue
\mciteSetBstMidEndSepPunct{\mcitedefaultmidpunct}
{\mcitedefaultendpunct}{\mcitedefaultseppunct}\relax
\EndOfBibitem
\bibitem[Frenkel and Warren(2015)Frenkel, and Warren]{Frenkel2015}
Frenkel,~D.; Warren,~P.~B. {{Gibbs, Boltzmann, and Negative Temperatures}}. \emph{Am. J. Phys.} \textbf{2015}, \emph{83}, 163--170\relax
\mciteBstWouldAddEndPuncttrue
\mciteSetBstMidEndSepPunct{\mcitedefaultmidpunct}
{\mcitedefaultendpunct}{\mcitedefaultseppunct}\relax
\EndOfBibitem
\bibitem[Huang(1987)]{Huang1987}
Huang,~K. \emph{{Statistical Mechanics}}, 2nd ed.; John Wiley \& Sons, Inc., 1987\relax
\mciteBstWouldAddEndPuncttrue
\mciteSetBstMidEndSepPunct{\mcitedefaultmidpunct}
{\mcitedefaultendpunct}{\mcitedefaultseppunct}\relax
\EndOfBibitem
\bibitem[Pathria and Beale(2011)Pathria, and Beale]{Pathria2011}
Pathria,~R.~K.; Beale,~P.~D. \emph{{Statistical Mechanics}}, 3rd ed.; Elsevier, 2011\relax
\mciteBstWouldAddEndPuncttrue
\mciteSetBstMidEndSepPunct{\mcitedefaultmidpunct}
{\mcitedefaultendpunct}{\mcitedefaultseppunct}\relax
\EndOfBibitem
\bibitem[Tuckerman(2010)]{tuckerman2010statistical}
Tuckerman,~M. \emph{{Statistical Mechanics: Theory and Molecular Simulation}}; Oxford university press, 2010\relax
\mciteBstWouldAddEndPuncttrue
\mciteSetBstMidEndSepPunct{\mcitedefaultmidpunct}
{\mcitedefaultendpunct}{\mcitedefaultseppunct}\relax
\EndOfBibitem
\bibitem[Swendsen(2012)]{Swendsen2012}
Swendsen,~R.~H. \emph{{An Introduction to Statistical Mechanics and Thermodynamics}}; Oxford University Press, 2012\relax
\mciteBstWouldAddEndPuncttrue
\mciteSetBstMidEndSepPunct{\mcitedefaultmidpunct}
{\mcitedefaultendpunct}{\mcitedefaultseppunct}\relax
\EndOfBibitem
\bibitem[Shell(2015)]{shell2015thermodynamics}
Shell,~M.~S. \emph{{Thermodynamics and Statistical Mechanics: An Integrated Approach}}; Cambridge University Press, 2015\relax
\mciteBstWouldAddEndPuncttrue
\mciteSetBstMidEndSepPunct{\mcitedefaultmidpunct}
{\mcitedefaultendpunct}{\mcitedefaultseppunct}\relax
\EndOfBibitem
\bibitem[Berdichevsky \latin{et~al.}(1991)Berdichevsky, Kunin, and Hussain]{Berdichevsky1991}
Berdichevsky,~V.; Kunin,~I.; Hussain,~F. Negative Temperature of Vortex Motion. \emph{Phys. Rev. A} \textbf{1991}, \emph{43}, 2050--2051\relax
\mciteBstWouldAddEndPuncttrue
\mciteSetBstMidEndSepPunct{\mcitedefaultmidpunct}
{\mcitedefaultendpunct}{\mcitedefaultseppunct}\relax
\EndOfBibitem
\bibitem[Dunkel and Hilbert(2006)Dunkel, and Hilbert]{Dunkel2006}
Dunkel,~J.; Hilbert,~S. {Phase Transitions in Small Systems: Microcanonical vs. Canonical Ensembles}. \emph{Phys. A: Stat. Mech.} \textbf{2006}, \emph{370}, 390--406\relax
\mciteBstWouldAddEndPuncttrue
\mciteSetBstMidEndSepPunct{\mcitedefaultmidpunct}
{\mcitedefaultendpunct}{\mcitedefaultseppunct}\relax
\EndOfBibitem
\bibitem[Schneider \latin{et~al.}(2014)Schneider, Mandt, Rapp, Braun, Weimer, Bloch, and Rosch]{Schneider2014}
Schneider,~U.; Mandt,~S.; Rapp,~A.; Braun,~S.; Weimer,~H.; Bloch,~I.; Rosch,~A. {Comment on ``Consistent Thermostatistics Forbids Negative Absolute Temperatures''}. \textbf{2014}, \relax
\mciteBstWouldAddEndPunctfalse
\mciteSetBstMidEndSepPunct{\mcitedefaultmidpunct}
{}{\mcitedefaultseppunct}\relax
\EndOfBibitem
\bibitem[Swendsen and Wang(2015)Swendsen, and Wang]{Swendsen2015}
Swendsen,~R.~H.; Wang,~J.-S. {{Gibbs Volume Entropy is Incorrect}}. \emph{Phys. Rev. E} \textbf{2015}, \emph{92}, 020103\relax
\mciteBstWouldAddEndPuncttrue
\mciteSetBstMidEndSepPunct{\mcitedefaultmidpunct}
{\mcitedefaultendpunct}{\mcitedefaultseppunct}\relax
\EndOfBibitem
\bibitem[Swendsen(2017)]{Swendsen2017}
Swendsen,~R. {Thermodynamics, Statistical Mechanics and Entropy}. \emph{Entropy} \textbf{2017}, \emph{19}, 603\relax
\mciteBstWouldAddEndPuncttrue
\mciteSetBstMidEndSepPunct{\mcitedefaultmidpunct}
{\mcitedefaultendpunct}{\mcitedefaultseppunct}\relax
\EndOfBibitem
\bibitem[Lavis(2019)]{Lavis2019}
Lavis,~D.~A. {The Question of Negative Temperatures in Thermodynamics and Statistical Mechanics}. \emph{Stud. Hist. Philos. Sci.} \textbf{2019}, \emph{67}, 26--63\relax
\mciteBstWouldAddEndPuncttrue
\mciteSetBstMidEndSepPunct{\mcitedefaultmidpunct}
{\mcitedefaultendpunct}{\mcitedefaultseppunct}\relax
\EndOfBibitem
\bibitem[Lustig(2019)]{Lustig2019}
Lustig,~R. Microcanonical Thermodynamics of Three and Four Atoms. \emph{J. Chem. Phys.} \textbf{2019}, \emph{150}, 074303\relax
\mciteBstWouldAddEndPuncttrue
\mciteSetBstMidEndSepPunct{\mcitedefaultmidpunct}
{\mcitedefaultendpunct}{\mcitedefaultseppunct}\relax
\EndOfBibitem
\bibitem[Barbatti(2022)]{Barbatti2022}
Barbatti,~M. Defining the Temperature of an Isolated Molecule. \emph{J. Chem. Phys.} \textbf{2022}, \emph{156}\relax
\mciteBstWouldAddEndPuncttrue
\mciteSetBstMidEndSepPunct{\mcitedefaultmidpunct}
{\mcitedefaultendpunct}{\mcitedefaultseppunct}\relax
\EndOfBibitem
\bibitem[Jensen and Shankar(1985)Jensen, and Shankar]{Jensen1985}
Jensen,~R.~V.; Shankar,~R. Statistical Behavior in Deterministic Quantum Systems with Few Degrees of Freedom. \emph{Phys. Rev. Lett.} \textbf{1985}, \emph{54}, 1879--1882\relax
\mciteBstWouldAddEndPuncttrue
\mciteSetBstMidEndSepPunct{\mcitedefaultmidpunct}
{\mcitedefaultendpunct}{\mcitedefaultseppunct}\relax
\EndOfBibitem
\bibitem[Chirikov \latin{et~al.}(1973)Chirikov, Izrailev, and Tayursky]{Chirikov1973}
Chirikov,~B.; Izrailev,~F.; Tayursky,~V. Numerical Experiments on the Statistical Behaviour of Dynamical Systems with a Few Degrees of Freedom. \emph{Comp. Phys. Commun.} \textbf{1973}, \emph{5}, 11--16\relax
\mciteBstWouldAddEndPuncttrue
\mciteSetBstMidEndSepPunct{\mcitedefaultmidpunct}
{\mcitedefaultendpunct}{\mcitedefaultseppunct}\relax
\EndOfBibitem
\bibitem[Cerino \latin{et~al.}(2014)Cerino, Gradenigo, Sarracino, Villamaina, and Vulpiani]{Cerino2014}
Cerino,~L.; Gradenigo,~G.; Sarracino,~A.; Villamaina,~D.; Vulpiani,~A. Fluctuations in Partitioning Systems with Few Degrees of Freedom. \emph{Phys. Rev. E} \textbf{2014}, \emph{89}, 042105\relax
\mciteBstWouldAddEndPuncttrue
\mciteSetBstMidEndSepPunct{\mcitedefaultmidpunct}
{\mcitedefaultendpunct}{\mcitedefaultseppunct}\relax
\EndOfBibitem
\bibitem[Lipka-Bartosik \latin{et~al.}(2023)Lipka-Bartosik, Perarnau-Llobet, and Brunner]{Lipka-Bartosik2023}
Lipka-Bartosik,~P.; Perarnau-Llobet,~M.; Brunner,~N. Operational Definition of the Temperature of a Quantum State. \emph{Phys. Rev. Lett.} \textbf{2023}, \emph{130}, 040401\relax
\mciteBstWouldAddEndPuncttrue
\mciteSetBstMidEndSepPunct{\mcitedefaultmidpunct}
{\mcitedefaultendpunct}{\mcitedefaultseppunct}\relax
\EndOfBibitem
\bibitem[Mitchison \latin{et~al.}(2022)Mitchison, Purkayastha, Brenes, Silva, and Goold]{Mitchison2022}
Mitchison,~M.~T.; Purkayastha,~A.; Brenes,~M.; Silva,~A.; Goold,~J. Taking the Temperature of a Pure Quantum State. \emph{Phys. Rev. A} \textbf{2022}, \emph{105}, L030201\relax
\mciteBstWouldAddEndPuncttrue
\mciteSetBstMidEndSepPunct{\mcitedefaultmidpunct}
{\mcitedefaultendpunct}{\mcitedefaultseppunct}\relax
\EndOfBibitem
\bibitem[Burke \latin{et~al.}(2023)Burke, Nakerst, and Haque]{Burke2023b}
Burke,~P.~C.; Nakerst,~G.; Haque,~M. Assigning Temperatures to Eigenstates. \emph{Phys. Rev. E} \textbf{2023}, \emph{107}, 024102\relax
\mciteBstWouldAddEndPuncttrue
\mciteSetBstMidEndSepPunct{\mcitedefaultmidpunct}
{\mcitedefaultendpunct}{\mcitedefaultseppunct}\relax
\EndOfBibitem
\bibitem[Phillies(2000)]{Phillies2000}
Phillies,~G.~D. \emph{Elementary Lectures in Statistical Mechanics}; Springer, New York, NY, 2000; pp 39--54\relax
\mciteBstWouldAddEndPuncttrue
\mciteSetBstMidEndSepPunct{\mcitedefaultmidpunct}
{\mcitedefaultendpunct}{\mcitedefaultseppunct}\relax
\EndOfBibitem
\bibitem[Santos \latin{et~al.}(2011)Santos, Polkovnikov, and Rigol]{Santos2011}
Santos,~L.~F.; Polkovnikov,~A.; Rigol,~M. Entropy of Isolated Quantum Systems after a Quench. \emph{Phys. Rev. Lett.} \textbf{2011}, \emph{107}, 040601\relax
\mciteBstWouldAddEndPuncttrue
\mciteSetBstMidEndSepPunct{\mcitedefaultmidpunct}
{\mcitedefaultendpunct}{\mcitedefaultseppunct}\relax
\EndOfBibitem
\bibitem[Andersen \latin{et~al.}(2001)Andersen, Bonderup, and Hansen]{Andersen2001}
Andersen,~J.~U.; Bonderup,~E.; Hansen,~K. On the Concept of Temperature for a Small Isolated System. \emph{J. Chem. Phys.} \textbf{2001}, \emph{114}, 6518--6525\relax
\mciteBstWouldAddEndPuncttrue
\mciteSetBstMidEndSepPunct{\mcitedefaultmidpunct}
{\mcitedefaultendpunct}{\mcitedefaultseppunct}\relax
\EndOfBibitem
\bibitem[Hartman \latin{et~al.}(2018)Hartman, Olsen, L{\"{u}}scher, Samani, Fallahi, Gardner, Manfra, and Folk]{Hartman2018}
Hartman,~N.; Olsen,~C.; L{\"{u}}scher,~S.; Samani,~M.; Fallahi,~S.; Gardner,~G.~C.; Manfra,~M.; Folk,~J. Direct Entropy Measurement in a Mesoscopic Quantum System. \emph{Nat. Phys.} \textbf{2018}, \emph{14}, 1083--1086\relax
\mciteBstWouldAddEndPuncttrue
\mciteSetBstMidEndSepPunct{\mcitedefaultmidpunct}
{\mcitedefaultendpunct}{\mcitedefaultseppunct}\relax
\EndOfBibitem
\bibitem[Puglisi \latin{et~al.}(2018)Puglisi, Sarracino, and Vulpiani]{Puglisi2018}
Puglisi,~A.; Sarracino,~A.; Vulpiani,~A. {Thermodynamics and Statistical Mechanics of Small Systems}. \emph{Entropy} \textbf{2018}, \emph{20}, 392\relax
\mciteBstWouldAddEndPuncttrue
\mciteSetBstMidEndSepPunct{\mcitedefaultmidpunct}
{\mcitedefaultendpunct}{\mcitedefaultseppunct}\relax
\EndOfBibitem
\bibitem[Chomaz and Gulminelli(2006)Chomaz, and Gulminelli]{Chomaz2006}
Chomaz,~P.; Gulminelli,~F. {The Challenges of Finite-System Statistical Mechanics}. \emph{European Physical Journal A} \textbf{2006}, \emph{30}, 317--331\relax
\mciteBstWouldAddEndPuncttrue
\mciteSetBstMidEndSepPunct{\mcitedefaultmidpunct}
{\mcitedefaultendpunct}{\mcitedefaultseppunct}\relax
\EndOfBibitem
\bibitem[Borderie and Frankland(2019)Borderie, and Frankland]{Borderie2019}
Borderie,~B.; Frankland,~J.~D. {Liquid–Gas Phase Transition in Nuclei}. \emph{Progress in Particle and Nuclear Physics} \textbf{2019}, \emph{105}, 82--138\relax
\mciteBstWouldAddEndPuncttrue
\mciteSetBstMidEndSepPunct{\mcitedefaultmidpunct}
{\mcitedefaultendpunct}{\mcitedefaultseppunct}\relax
\EndOfBibitem
\bibitem[Gross(2006)]{Gross2006}
Gross,~D.~H. {Reconciliation of Statistical Mechanics and Astro-Physical Statistics: The Errors of Conventional Canonical Thermostatistics}. \emph{Comptes Rendus Physique} \textbf{2006}, \emph{7}, 311--317\relax
\mciteBstWouldAddEndPuncttrue
\mciteSetBstMidEndSepPunct{\mcitedefaultmidpunct}
{\mcitedefaultendpunct}{\mcitedefaultseppunct}\relax
\EndOfBibitem
\bibitem[Puglisi \latin{et~al.}(2017)Puglisi, Sarracino, and Vulpiani]{Puglisi2017}
Puglisi,~A.; Sarracino,~A.; Vulpiani,~A. {Temperature In and Out of Equilibrium: A Review of Concepts, Tools and Attempts}. \emph{Physics Reports} \textbf{2017}, \emph{709-710}, 1--60\relax
\mciteBstWouldAddEndPuncttrue
\mciteSetBstMidEndSepPunct{\mcitedefaultmidpunct}
{\mcitedefaultendpunct}{\mcitedefaultseppunct}\relax
\EndOfBibitem
\bibitem[Tolman(1938)]{Tolman1938}
Tolman,~R.~C. \emph{{The Principles of Statistical Mechanics}}; Oxford at the Clarendon Press, 1938\relax
\mciteBstWouldAddEndPuncttrue
\mciteSetBstMidEndSepPunct{\mcitedefaultmidpunct}
{\mcitedefaultendpunct}{\mcitedefaultseppunct}\relax
\EndOfBibitem
\bibitem[Mayer and Mayer(1940)Mayer, and Mayer]{Mayer1940}
Mayer,~J.~E.; Mayer,~M.~G. \emph{{Statistical Mechanics}}; John Wiley \& Sons, Inc.: New York, NY, 1940\relax
\mciteBstWouldAddEndPuncttrue
\mciteSetBstMidEndSepPunct{\mcitedefaultmidpunct}
{\mcitedefaultendpunct}{\mcitedefaultseppunct}\relax
\EndOfBibitem
\bibitem[Landau and Lifshitz(2013)Landau, and Lifshitz]{landau2013statistical}
Landau,~L.~D.; Lifshitz,~E.~M. \emph{Statistical Physics: Volume 5}; Elsevier, 2013; Vol.~5\relax
\mciteBstWouldAddEndPuncttrue
\mciteSetBstMidEndSepPunct{\mcitedefaultmidpunct}
{\mcitedefaultendpunct}{\mcitedefaultseppunct}\relax
\EndOfBibitem
\bibitem[Park \latin{et~al.}(2022)Park, Kim, and Yi]{Park2022}
Park,~H.; Kim,~Y.~W.; Yi,~J. Entropies of the Microcanonical Ensemble. \emph{AIP Adv.} \textbf{2022}, \emph{12}\relax
\mciteBstWouldAddEndPuncttrue
\mciteSetBstMidEndSepPunct{\mcitedefaultmidpunct}
{\mcitedefaultendpunct}{\mcitedefaultseppunct}\relax
\EndOfBibitem
\bibitem[Burke and Haque(2023)Burke, and Haque]{Burke2023}
Burke,~P.~C.; Haque,~M. Entropy and Temperature in Finite Isolated Quantum Systems. \emph{Phys. Rev. E} \textbf{2023}, \emph{107}, 034125\relax
\mciteBstWouldAddEndPuncttrue
\mciteSetBstMidEndSepPunct{\mcitedefaultmidpunct}
{\mcitedefaultendpunct}{\mcitedefaultseppunct}\relax
\EndOfBibitem
\bibitem[Gurarie(2007)]{Gurarie2007}
Gurarie,~V. {The Equivalence Between the Canonical and Microcanonical Ensembles When Applied to Large Systems}. \emph{Am. J. Phys.} \textbf{2007}, \emph{75}, 747--751\relax
\mciteBstWouldAddEndPuncttrue
\mciteSetBstMidEndSepPunct{\mcitedefaultmidpunct}
{\mcitedefaultendpunct}{\mcitedefaultseppunct}\relax
\EndOfBibitem
\bibitem[Srednicki(1994)]{Srednicki1994}
Srednicki,~M. {Chaos and quantum thermalization}. \emph{Physical Review E} \textbf{1994}, \emph{50}, 888--901\relax
\mciteBstWouldAddEndPuncttrue
\mciteSetBstMidEndSepPunct{\mcitedefaultmidpunct}
{\mcitedefaultendpunct}{\mcitedefaultseppunct}\relax
\EndOfBibitem
\bibitem[Lezama \latin{et~al.}(2021)Lezama, Torres-Herrera, P{\'{e}}rez-Bernal, {Bar Lev}, and Santos]{Lezama2021}
Lezama,~T. L.~M.; Torres-Herrera,~E.~J.; P{\'{e}}rez-Bernal,~F.; {Bar Lev},~Y.; Santos,~L.~F. {Equilibration time in many-body quantum systems}. \emph{Physical Review B} \textbf{2021}, \emph{104}, 085117\relax
\mciteBstWouldAddEndPuncttrue
\mciteSetBstMidEndSepPunct{\mcitedefaultmidpunct}
{\mcitedefaultendpunct}{\mcitedefaultseppunct}\relax
\EndOfBibitem
\bibitem[Rigol \latin{et~al.}(2008)Rigol, Dunjko, and Olshanii]{Rigol2008}
Rigol,~M.; Dunjko,~V.; Olshanii,~M. {Thermalization and its mechanism for generic isolated quantum systems}. \emph{Nature} \textbf{2008}, \emph{452}, 854--858\relax
\mciteBstWouldAddEndPuncttrue
\mciteSetBstMidEndSepPunct{\mcitedefaultmidpunct}
{\mcitedefaultendpunct}{\mcitedefaultseppunct}\relax
\EndOfBibitem
\bibitem[Hill(1994)]{hill1994thermodynamics}
Hill,~T.~L. \emph{Thermodynamics of Small Systems}; Courier Corporation, 1994\relax
\mciteBstWouldAddEndPuncttrue
\mciteSetBstMidEndSepPunct{\mcitedefaultmidpunct}
{\mcitedefaultendpunct}{\mcitedefaultseppunct}\relax
\EndOfBibitem
\bibitem[Griffiths and Schroeter(2018)Griffiths, and Schroeter]{griffiths2018introduction}
Griffiths,~D.~J.; Schroeter,~D.~F. \emph{Introduction to Quantum Mechanics}; Cambridge University Press, 2018\relax
\mciteBstWouldAddEndPuncttrue
\mciteSetBstMidEndSepPunct{\mcitedefaultmidpunct}
{\mcitedefaultendpunct}{\mcitedefaultseppunct}\relax
\EndOfBibitem
\bibitem[Miranda and Bertoldi(2013)Miranda, and Bertoldi]{Miranda2013}
Miranda,~E.~N.; Bertoldi,~D.~S. Thermostatistics of Small Systems: Exact Results in the Microcanonical Formalism. \emph{Eur. J. Phys.} \textbf{2013}, \emph{34}, 1075--1087\relax
\mciteBstWouldAddEndPuncttrue
\mciteSetBstMidEndSepPunct{\mcitedefaultmidpunct}
{\mcitedefaultendpunct}{\mcitedefaultseppunct}\relax
\EndOfBibitem
\bibitem[Shirts(2021)]{Shirts2021}
Shirts,~R.~B. A Comparison of Boltzmann and Gibbs Definitions of Microcanonical Entropy for Small Systems. \emph{AIP Adv.} \textbf{2021}, \emph{11}\relax
\mciteBstWouldAddEndPuncttrue
\mciteSetBstMidEndSepPunct{\mcitedefaultmidpunct}
{\mcitedefaultendpunct}{\mcitedefaultseppunct}\relax
\EndOfBibitem
\bibitem[Kardar(2007)]{kardar2007statistical}
Kardar,~M. \emph{Statistical Physics of Particles}; Cambridge University Press, 2007\relax
\mciteBstWouldAddEndPuncttrue
\mciteSetBstMidEndSepPunct{\mcitedefaultmidpunct}
{\mcitedefaultendpunct}{\mcitedefaultseppunct}\relax
\EndOfBibitem
\bibitem[Ramsey(1956)]{Ramsey1956}
Ramsey,~N.~F. {Thermodynamics and Statistical Mechanics at Negative Absolute Temperatures}. \emph{Phys. Rev.} \textbf{1956}, \emph{103}, 20--28\relax
\mciteBstWouldAddEndPuncttrue
\mciteSetBstMidEndSepPunct{\mcitedefaultmidpunct}
{\mcitedefaultendpunct}{\mcitedefaultseppunct}\relax
\EndOfBibitem
\bibitem[Vilar and Rubi(2014)Vilar, and Rubi]{Vilar2014}
Vilar,~J. M.~G.; Rubi,~J.~M. {Communication: System-Size Scaling of Boltzmann and Alternate Gibbs Entropies}. \emph{J. Chem. Phys.} \textbf{2014}, \emph{140}, 201101\relax
\mciteBstWouldAddEndPuncttrue
\mciteSetBstMidEndSepPunct{\mcitedefaultmidpunct}
{\mcitedefaultendpunct}{\mcitedefaultseppunct}\relax
\EndOfBibitem
\bibitem[Purcell and Pound(1951)Purcell, and Pound]{Purcell1951}
Purcell,~E.~M.; Pound,~R.~V. {A Nuclear Spin System at Negative Temperature}. \emph{Phys. Rev.} \textbf{1951}, \emph{81}, 279--280\relax
\mciteBstWouldAddEndPuncttrue
\mciteSetBstMidEndSepPunct{\mcitedefaultmidpunct}
{\mcitedefaultendpunct}{\mcitedefaultseppunct}\relax
\EndOfBibitem
\bibitem[Javan(1959)]{Javan1959}
Javan,~A. {Possibility of Production of Negative Temperature in Gas Discharges}. \emph{Phys. Rev. Lett.} \textbf{1959}, \emph{3}, 87--89\relax
\mciteBstWouldAddEndPuncttrue
\mciteSetBstMidEndSepPunct{\mcitedefaultmidpunct}
{\mcitedefaultendpunct}{\mcitedefaultseppunct}\relax
\EndOfBibitem
\bibitem[Hakonen and Lounasmaa(1994)Hakonen, and Lounasmaa]{Hakonen1994}
Hakonen,~P.; Lounasmaa,~O.~V. {Negative Absolute Temperatures: "Hot" Spins in Spontaneous Magnetic Order}. \emph{Science} \textbf{1994}, \emph{265}, 1821--1825\relax
\mciteBstWouldAddEndPuncttrue
\mciteSetBstMidEndSepPunct{\mcitedefaultmidpunct}
{\mcitedefaultendpunct}{\mcitedefaultseppunct}\relax
\EndOfBibitem
\bibitem[Mosk(2005)]{Mosk2005}
Mosk,~A.~P. {Atomic Gases at Negative Kinetic Temperature}. \emph{Phys. Rev. Lett.} \textbf{2005}, \emph{95}, 040403\relax
\mciteBstWouldAddEndPuncttrue
\mciteSetBstMidEndSepPunct{\mcitedefaultmidpunct}
{\mcitedefaultendpunct}{\mcitedefaultseppunct}\relax
\EndOfBibitem
\bibitem[Rapp \latin{et~al.}(2010)Rapp, Mandt, and Rosch]{Rapp2010}
Rapp,~A.; Mandt,~S.; Rosch,~A. {Equilibration Rates and Negative Absolute Temperatures for Ultracold Atoms in Optical Lattices}. \emph{Phys. Rev. Lett.} \textbf{2010}, \emph{105}, 220405\relax
\mciteBstWouldAddEndPuncttrue
\mciteSetBstMidEndSepPunct{\mcitedefaultmidpunct}
{\mcitedefaultendpunct}{\mcitedefaultseppunct}\relax
\EndOfBibitem
\bibitem[Braun \latin{et~al.}(2013)Braun, Ronzheimer, Schreiber, Hodgman, Rom, Bloch, and Schneider]{Braun2013}
Braun,~S.; Ronzheimer,~J.~P.; Schreiber,~M.; Hodgman,~S.~S.; Rom,~T.; Bloch,~I.; Schneider,~U. {Negative Absolute Temperature for Motional Degrees of Freedom}. \emph{Science} \textbf{2013}, \emph{339}, 52--55\relax
\mciteBstWouldAddEndPuncttrue
\mciteSetBstMidEndSepPunct{\mcitedefaultmidpunct}
{\mcitedefaultendpunct}{\mcitedefaultseppunct}\relax
\EndOfBibitem
\bibitem[Baudin \latin{et~al.}(2023)Baudin, Garnier, Fusaro, Berti, Michel, Krupa, Millot, and Picozzi]{Baudin2023}
Baudin,~K.; Garnier,~J.; Fusaro,~A.; Berti,~N.; Michel,~C.; Krupa,~K.; Millot,~G.; Picozzi,~A. {Observation of Light Thermalization to Negative-Temperature Rayleigh-Jeans Equilibrium States in Multimode Optical Fibers}. \emph{Phys. Rev. Lett.} \textbf{2023}, \emph{130}, 063801\relax
\mciteBstWouldAddEndPuncttrue
\mciteSetBstMidEndSepPunct{\mcitedefaultmidpunct}
{\mcitedefaultendpunct}{\mcitedefaultseppunct}\relax
\EndOfBibitem
\bibitem[Abraham and Penrose(2017)Abraham, and Penrose]{Abraham2017}
Abraham,~E.; Penrose,~O. {Physics of Negative Absolute Temperatures}. \emph{Phys. Rev. E} \textbf{2017}, \emph{95}, 012125\relax
\mciteBstWouldAddEndPuncttrue
\mciteSetBstMidEndSepPunct{\mcitedefaultmidpunct}
{\mcitedefaultendpunct}{\mcitedefaultseppunct}\relax
\EndOfBibitem
\bibitem[Baldovin \latin{et~al.}(2021)Baldovin, Iubini, Livi, and Vulpiani]{Baldovin2021}
Baldovin,~M.; Iubini,~S.; Livi,~R.; Vulpiani,~A. {Statistical Mechanics of Systems with Negative Temperature}. \emph{Physics Reports} \textbf{2021}, \emph{923}, 1--50\relax
\mciteBstWouldAddEndPuncttrue
\mciteSetBstMidEndSepPunct{\mcitedefaultmidpunct}
{\mcitedefaultendpunct}{\mcitedefaultseppunct}\relax
\EndOfBibitem
\end{mcitethebibliography}
\bibliographystyle{achemso}
\end{document}


\tableofcontents

\addcontentsline{toc}{section}{S1. Derivation of the energy expression for a quantum harmonic oscillator in the microcanonical ensemble valid at all temperatures}
\section{S1. Derivation of the energy expression for a quantum harmonic oscillator in the microcanonical ensemble valid at all temperatures}
Recall from the main text that for a quantum harmonic oscillator,
\begin{equation*}
E_n=h \nu\left(n+\frac{1}{2}\right)
\end{equation*}
where $n=0,1,2,...$ and $\nu$ represents the frequency of vibration of the oscillator. It follows, as mentioned in the main text, that the phase space volume is
\begin{eqnarray*}
    \Omega=\frac{E}{h\nu}+\frac{1}{2}
\end{eqnarray*}
Further, the phase space density is
\begin{eqnarray*}
    \omega = \frac{\epsilon}{h\nu}\\
\end{eqnarray*}
Finally, the phase space density with a relative energy constraint is
\begin{eqnarray*}
        \omega_r = \frac{\epsilon_r E}{h\nu}
\end{eqnarray*}
At high $T$, the energy of the system changes smoothly, and thus we obtain the results shown in Table 1 of the main text, by defining the temperature as
\begin{equation*}
    T=\frac{\partial E}{\partial S}
\end{equation*}

Since the energy of the system changes only discretely at low $T$, we define the temperature as
\begin{equation*}
    T=\frac{\Delta E}{\Delta S}
\end{equation*}
In this case, we also need to define the Boltzmann partition functions based on discrete jumps in states as
\begin{equation*}
    \omega=\frac{\Delta \Omega}{\Delta E}\epsilon = \frac{\epsilon}{h\nu}
\end{equation*}
and
\begin{equation*}
    \omega_r=\frac{\Delta \Omega}{\Delta E}\epsilon_r E= \frac{\epsilon_r E}{h\nu}
\end{equation*}
Now, considering the adjacent system states with energy $E$ and $E+h\nu$, we obtain
\begin{equation*}
    T=\frac{h\nu}{S(E+h\nu-E_0)-S(E-E_0)}
\end{equation*}
where $E_0=\frac{1}{2}h\nu$ is the ground-state energy of the quantum harmonic oscillator. For the Gibbs entropy, we obtain
\begin{equation}
    T_G=\frac{h\nu}{k_B\ln\left(\frac{\frac{E-E_0+h\nu}{h\nu}+\frac{1}{2}}{\frac{E-E_0}{h\nu}+\frac{1}{2}}\right)}
\end{equation}
Inverting to solve for the energy, we get
\begin{equation}
    \frac{E-E_0}{h\nu} =  \frac{3-\exp\left(\frac{h\nu}{k_BT_G}\right)}{2\exp\left(\frac{h\nu}{k_BT_G}\right)-2}
\end{equation}
or,
\begin{equation}
    E = \frac{1}{2}h\nu + h\nu\left(\frac{3-\exp\left(\frac{h\nu}{k_BT_G}\right)}{2\exp\left(\frac{h\nu}{k_BT_G}\right)-2}\right)
\end{equation}
Now, considering the original Boltzmann entropy, we obtain
\begin{equation}
    T_B = \frac{h\nu}{0} \rightarrow \infty
\end{equation}
Finally, the consideration of the relative-energy-constrained Boltzmann entropy leads to
\begin{equation}
    T_{B_r} = \frac{h\nu}{k_B\ln\left(\frac{E-E_0+h\nu}{E-E_0}\right)}
\end{equation}
so that
\begin{eqnarray*}
    E  = E_0 + \frac{h\nu}{\exp\left(\frac{h\nu}{k_BT_{B_r}}\right)-1}
\end{eqnarray*}
Now, recall the energy of the quantum harmonic oscillator in the canonical ensemble. It is
\begin{eqnarray*}
        E = \frac{1}{2}h\nu +  \frac{h\nu}{\exp\left(\frac{h\nu}{k_BT}\right)-1}
\end{eqnarray*}
Thus, \textit{the relative-energy constrained Boltzmann entropy} is able to match the canonical prediction for energy, whereas both the original Boltzmann and the Gibbs entropy fail! Further, note that the Gibbs entropy expression for energy can be written as
\begin{equation}
    E =  \frac{1}{2}h\nu +  \left(\frac{h\nu}{\exp\left(\frac{h\nu}{k_BT_G}\right)-1} - \frac{1}{2}h\nu\right)
\end{equation}
or,
\begin{equation}
    E =  \frac{h\nu}{\exp\left(\frac{h\nu}{k_BT_G}\right)-1}
\end{equation}
Thus, at $T=0$, the Gibbs entropy predicts an energy of $0$, which is incorrect, as it fails to recover the zero-point energy.

\addcontentsline{toc}{section}{S2. High-temperature limits for the energy of a quantum harmonic oscillator}
\section{S2. High-temperature limits for the energy of a quantum harmonic oscillator}
For large $T$, we can use the Taylor series expansion
\begin{equation*}
    \exp\left(\frac{h\nu}{k_BT}\right) \approx 1 + \frac{h\nu}{k_BT}
\end{equation*}
This leads to, for the Gibbs entropy,
\begin{eqnarray*}
    E = h\nu \left(\frac{2-\frac{h\nu}{k_BT_G}}{2 \frac{h\nu}{k_BT_G}}\right) = k_BT_G - \frac{h\nu}{2}
\end{eqnarray*}
as reported in the main text for high temperature.

Analogously, for the original Boltzmann entropy, since $T$ is always infinite, the high-temperature limit also leads to
\begin{eqnarray*}
   T_B \rightarrow \infty
\end{eqnarray*}
Finally, the relative-energy-constrained Boltzmann entropy leads to
\begin{eqnarray*}
    E = \frac{h\nu}{\frac{h\nu}{k_BT_r}} = k_BT_r
\end{eqnarray*}
also as reported in the main text.

\addcontentsline{toc}{section}{S3. Examination of the energy-dependent energy width proposed in previous work}
\section{S3. Examination of the energy-dependent energy width proposed in previous work}
As discussed in the main text, the concept of an energy-dependent energy tolerance has received some consideration previously.\cite{Gurarie2007,Burke2023} The energy tolerance proposed in ref. \cite{Burke2023} is
\begin{equation*}
    \epsilon = \alpha^{-1} \sqrt{2\pi k_B T_C^2 C_C}
\end{equation*}
where $\alpha$ is a constant (proposed to be exactly 1 in ref. \cite{Gurarie2007}), $k_B$ is the Boltzmann constant, $T_C$ is the temperature of the system in the canonical ensemble, and $C_C$ is the heat capacity of the system in the canonical ensemble, defined as $C_C=\frac{\partial E}{\partial T_C}$, with $E$ denoting the system's energy. Considering the example of the $d$-dimensional monatomic ideal gas discussed in the main text, recall that
\begin{eqnarray*}
    \omega = \frac{V^N}{h^{dN}N!}\frac{(2\pi m)^{\frac{dN}{2}}}{\Gamma\left(\frac{dN}{2}\right)}E^{\frac{dN}{2}-1}\epsilon
\end{eqnarray*}
where $V$ is the volume of the system, $h$ the Planck's constant, and $N$ the number of atoms in the system. For a $d$-dimensional ideal gas, the canonical ensemble can be used to show that\cite{kaznessis2011statistical} 
\begin{eqnarray*}
    E = \frac{d}{2}Nk_BT_C
\end{eqnarray*}
so that
\begin{eqnarray*}
    C_C = \frac{d}{2}Nk_B = \frac{E}{T_C}
\end{eqnarray*}
Thus, we have
\begin{eqnarray*}
    \epsilon = \alpha^{-1} \sqrt{2\pi k_B T_C E} =2\alpha^{-1}E \sqrt{  \frac{\pi}{dN}} 
\end{eqnarray*}
Using this expression for $\epsilon$, we obtain
\begin{eqnarray*}
    \omega = 2\alpha^{-1} \sqrt{  \frac{\pi}{dN}} \frac{V^N}{h^{dN}N!}\frac{(2\pi m)^{\frac{dN}{2}}}{\Gamma\left(\frac{dN}{2}\right)}E^{\frac{dN}{2}}  
\end{eqnarray*}
This gives the Boltzmann temperature using $S_B=k_B\ln\omega$ and $T_B=\frac{\partial E}{\partial S_B}$ to be
\begin{eqnarray*}
    E = \frac{dN}{2}k_BT
\end{eqnarray*}
thus seeming to resolve the challenges with the Boltzmann entropy. 
However, problems arise when one examines the case of a low-temperature quantum harmonic oscillator. Here, 
\begin{eqnarray*}
        E_C  = E_0 + \frac{h\nu}{\exp\left(\frac{h\nu}{k_BT_{C}}\right)-1}
\end{eqnarray*}
so that 
\begin{eqnarray*}
    C_C = \frac{h^2\nu^2 \exp\left(\frac{h\nu}{k_BT_{C}}\right)}{k_BT_C^2\left(\exp\left(\frac{h\nu}{k_BT_{C}}\right)-1\right)^2}
\end{eqnarray*}
and
\begin{eqnarray*}
    k_B T_C^2 C_C = \frac{h^2\nu^2 \exp\left(\frac{h\nu}{k_BT_{C}}\right)}{\left(\exp\left(\frac{h\nu}{k_BT_{C}}\right)-1\right)^2}
\end{eqnarray*}
Thus, the energy-dependent tolerance proposed in refs. \cite{Gurarie2007} and \cite{Burke2023} leads to
\begin{eqnarray*}
    \epsilon = \alpha^{-1} \sqrt{2\pi\exp\left(\frac{h\nu}{k_BT_{C}}\right) \frac{h^2\nu^2 }{\left(\exp\left(\frac{h\nu}{k_BT_{C}}\right)-1\right)^2}}    \end{eqnarray*}
i.e.,
\begin{equation}
\epsilon = \alpha^{-1} \sqrt{2\pi  \left(\frac{h\nu}{E-E_0} + 1\right) (E-E_0)^2}    
\end{equation}
Hence, we obtain the number of states as
\begin{equation}
\omega = \frac{\alpha^{-1}}{h\nu} \sqrt{2\pi  \left(\frac{h\nu}{E-E_0} + 1\right) (E-E_0)^2}    
\end{equation}
Therefore, we have
\begin{equation*}
    T=\frac{h\nu}{S(E+h\nu-E_0)-S(E-E_0)} = \frac{h\nu}{k_B\ln \left\lbrace \sqrt{\frac{\frac{h\nu}{E+h\nu-E_0} + 1}{\frac{h\nu}{E-E_0} + 1}}\left(\frac{E+h\nu-E_0}{E-E_0}\right)\right\rbrace}
\end{equation*}
an equation that does not agree with the equation of state for a quantum harmonic oscillator. Therefore, the energy width expression proposed in refs. \cite{Gurarie2007} and \cite{Burke2023} does not lead to the correct thermodynamic behavior of an isolated quantum harmonic oscillator, unlike the entropy expression proposed in this work.

\addcontentsline{toc}{section}{S4. Canonical formulation of the two-level system}
\section{S4. Canonical formulation of the two-level system}
The canonical ensemble formulation of the two-level system discussed in the main text is a standard problem, whose solution is presented here for convenience. In this case, the partition function is given as:
\begin{eqnarray*}
    Q(N,V,T) = \sum_i e^{-\beta E_i}
\end{eqnarray*}
where $\beta=\frac{1}{k_BT}$ and $E_i$ is the energy of the $i$\textsuperscript{th} microstate. Thus, we have
\begin{eqnarray*}
    Q(N,V,T) = {N \choose 0}e^{-\beta\times 0E_1} + {N \choose 1} e^{-\beta\times 1E_1} + {N \choose 2} e^{-\beta\times 2E_1} +  \hdots + {N \choose N} e^{-\beta\times NE_1}
\end{eqnarray*}
which can be simplified to
\begin{eqnarray*}
    Q(N,V,T) = (1+\exp(-\beta E_1))^N
\end{eqnarray*}
In the canonical ensemble, the Helmholtz free energy of the system can be obtained as
\begin{eqnarray*}
    A=-k_B T \ln Q = -Nk_B T \ln \left(1+e^{-\frac{E_1}{k_B T}}\right)
\end{eqnarray*}
so that
\begin{eqnarray*}
    S=- \frac{\partial A}{\partial T} = Nk_B \left[ \ln \left(1+e^{-\frac{E_1}{k_B T}}\right) + \frac{E_1 e^{-\frac{E_1}{k_BT}}}{k_B T \left(1+e^{-\frac{E_1}{k_B T}}\right)} \right]
\end{eqnarray*}
Therefore, the internal energy of the system can be calculated to be
\begin{eqnarray*}
    E=A+TS = \frac{NE_1 e^{-\frac{E_1}{k_BT}}}{1+e^{-\frac{E_1}{k_BT}}}
\end{eqnarray*}
In the main text, the non-dimensional temperature of the two-level system, i.e., $k_B T/E_1$ is plotted versus the non-dimensional energy, i.e., $E/E_1$. Thus, we invert the above equation to express $T$ in terms of $E$ to obtain:
\begin{eqnarray*}
    \frac{k_BT_C}{E_1} = \frac{1}{\ln\left(\frac{N-\frac{E}{E_1}}{\frac{E}{E_1}}\right)}
\end{eqnarray*}
where we have replaced $T$ with $T_C$ to indicate that the result was obtained in the canonical ensemble. This equation is plotted as the canonical ensemble result in Figure 4 of the main text. Although this equation is identical to the large-system-size limit for both $T_B$ and $T_{B_r}$ as presented in the main text, it is instructive to find out which expression---$T_B$ or $T_{B_r}$---is closer to $T_C$. An examination of Figure 4 of the main text, taking into account the staggered nature of entropy calculations due to the use of finite differences, reveals that it is $T_{B_r}$!

\addcontentsline{toc}{section}{S5. Examining the consistency condition mentioned in previous work}
\section{S5. Examining the consistency condition mentioned in previous work}
Based on the fundamental thermodynamic relation
\begin{eqnarray*}
  dS = \left(\frac{\partial S}{\partial E}\right)dE + \left(\frac{\partial S}{\partial V}\right)dV + \sum_i \left(\frac{\partial S}{\partial A_i}\right)dA_i \\ = \frac{1}{T}dE + \frac{P}{T}dV + \sum_i \frac{a_i}{T}dA_i
\end{eqnarray*}
where $A_i$ is an extensive thermodynamic variable, $a_i$ its intensive conjugate variable, and $P$ the pressure of the system, Dunkel and Hilbert \cite{Dunkel2014} stated that under adiabatic conditions, it can be shown that 
\begin{equation*}
    T\left(\frac{\partial S}{\partial A_j}\right)_E = -\left\langle \frac{\partial H}{\partial A_j}\right\rangle
\end{equation*}
where $A_j\in \{V,A_i\}$. Further, they mentioned that the equation
\begin{eqnarray}
\label{eq:adiab1}
T_G \frac{\partial S_G}{\partial A_j} = \frac{1}{\omega} \frac{\partial}{\partial A_j} \text{Tr}[\Theta(E-H)]\\
\label{eq:adiab2}
=-\text{Tr}\left[\frac{\partial H}{\partial A_j}\frac{\delta(E-H)}{\omega}\right]=-\left\langle\frac{\partial H}{\partial A_j}\right\rangle
\end{eqnarray}
proves that the Gibbs entropy is thermodynamically consistent. In contrast, however, the requirement of this condition for a valid entropy definition has been contested \cite{Buonsante2016}. Moreover, we show here that the proposed condition is not satisfied even by the Gibbs entropy. Indeed, upon closely examining the derivation, one sees that Eq. (\ref{eq:adiab1}) relies upon the validity of Gibbs entropy (phase space volume definition), whereas Eq. (\ref{eq:adiab2}) relies on the validity of the Boltzmann entropy (phase space density definition). It follows that the proof of thermodynamic consistency demonstrated in ref. \cite{Dunkel2014} relies on the mixing of two entropy definitions, which is incorrect \cite{uline2008generalized,corti2023microcanonical}. For this reason, Gibbs entropy itself does not satisfy Eqs. (\ref{eq:adiab1})-(\ref{eq:adiab2}), and therefore, this argument cannot be used to support it.  

\bibliography{apssamp}